\documentclass[12pt]{article}
\usepackage{graphics,graphicx}
\usepackage{amssymb,epsfig,amsmath,euscript,array}
\usepackage[nosort]{cite}
\usepackage{axodraw4j}
\usepackage{pstricks}
\usepackage{color}
\usepackage{bbm}

\makeatletter
\@addtoreset{equation}{section}
\makeatother

\newcounter{multieqs}

\newcommand{\be}{\begin{equation}}
\newcommand{\ee}{\end{equation}}

\newcommand{\bm}[1]{\mbox{\boldmath $#1$}}

\newcommand{\kslash}{k \!\!\! / }

\newcommand{\lslash}{l \!\! / }
\newcommand{\Pslash}{P \!\!\!\! / }

\newcommand{\islash}{i \!\!\! / }
\newcommand{\jslash}{j \!\!\! / }
\newcommand{\aslash}{a \!\!\! / }
\newcommand{\bslash}{{b \hspace{-6pt} \slash} }

\newcommand{\onslash}{1 \!\!\! / }
\newcommand{\twslash}{2 \!\!\!/ }
\newcommand{\thslash}{3 \!\!\!/ }
\newcommand{\foslash}{4 \!\!\! / }
\newcommand{\fislash}{5 \!\!\! / }

\newcommand{\mslash}{m \!\!\! / }

\def\bd{\begin{document}}
\def\ed{\end{document}}
\def\nn{\nonumber}
\def\bea{\begin{eqnarray}}
\def\eea{\end{eqnarray}}

\def\ab{(ijab)}
\def\ba{(ijba)}
\def\ijab{{\tr}_{+}(\islash\, \jslash\, \aslash \, \bslash)}
\def\ijba{{\tr}_{+}(\islash\, \jslash\, \bslash \, \aslash)}
\def\ijaP{{\tr}_{+}(\islash\, \jslash\, \aslash \, \Pslash)}
\def\ijPLa{{\tr}_{+}(\islash\, \jslash\, \Pslash_L \, \aslash)}
\def\ijaPL{{\tr}_{+}(\islash\, \jslash\, \aslash \, \Pslash_L)}
\def\ijPLza{{\tr}_{+}(\islash\, \jslash\, \Pslash_{L;z} \, \aslash)}
\def\ijaPLz{{\tr}_{+}(\islash\, \jslash\, \aslash \, \Pslash_{L;z})}
\def\ijPa{{\tr}_{+}(\islash\, \jslash\, \Pslash \, \aslash)}
\def\iaPb{{\tr}_{+}(\islash\, \aslash\, \Pslash \, \bslash)}
\def\ibPa{{\tr}_{+}(\islash\, \bslash\, \Pslash \, \aslash)}
\def\ijPmu{{\tr}_{+}(\islash\, \jslash\, \Pslash \, \mu)}
\def\ibmuP{{\tr}_{+}(\islash\, \bslash\, \mu \, \Pslash)}
\def\ibmua{{\tr}_{+}(\islash\, \bslash\, \mu \, \aslash)}
\def\iamub{{\tr}_{+}(\islash\, \aslash\, \mu \, \bslash)}
\def\jaPb{{\tr}_{+}(\jslash\, \aslash\, \Pslash \, \bslash)}
\def\ijmuP{{\tr}_{+}(\islash\, \jslash\, \mu \, \Pslash)}
\def\ijmum{{\tr}_{+}(\islash\, \jslash\, \mu \, \mslash)}
\def\ijmmu{{\tr}_{+}(\islash\, \jslash\, \mslash \, \mu)}
\def\ijmP{{\tr}_{+}(\islash\, \jslash\, \mslash \, \Pslash)}
\def\iabP{{\tr}_{+}(\islash\, \aslash\, \bslash \, \Pslash)}
\def\ijbP{{\tr}_{+}(\islash\, \jslash\, \bslash \, \Pslash)}
\def\jbPa{{\tr}_{+}(\jslash\, \bslash\, \Pslash \, \aslash)}
\def\ijPb{{\tr}_{+}(\islash\, \jslash\, \Pslash \, \bslash)}
\def\jbmua{{\tr}_{+}(\jslash\, \bslash\, \mu \, \aslash)}

\def\loablt{ {\tr}_{+}(\lslash_1\, \aslash \, \bslash\, \lslash_2)}

\def\ijlolt{{\tr}_{+}(\islash\, \jslash\, \lslash_1 \, \lslash_2)}
\def\ijltlo{{\tr}_{+}(\islash\, \jslash\, \lslash_2 \, \lslash_1)}
\def\ibloa{{\tr}_{+}(\islash\, \bslash\, \lslash_1 \, \aslash)}
\def\jaltb{{\tr}_{+}(\jslash\, \aslash\, \lslash_2 \, \bslash)}
\def\ialtb{{\tr}_{+}(\islash\, \aslash\, \lslash_2 \, \bslash)}
\def\bltloa{{\tr}_{+}(\bslash\, \lslash_2\, \lslash_1 \, \aslash)}
\def\jbloa{{\tr}_{+}(\jslash\, \bslash\, \lslash_1 \, \aslash)}
\def\ibPb{{\tr}_{+}(\islash\, \bslash\, \Pslash \, \bslash)}
\def\ijltb{{\tr}_{+}(\islash\, \jslash\, \lslash_2 \, \bslash)}

\def\ijloa{{\tr}_{+}(\islash\, \jslash\,  \lslash_1 \, \aslash)}
\def\ijblt{{\tr}_{+}(\islash\, \jslash\,  \bslash \, \lslash_2)}

\def\jakb{{\tr}_{+}(\jslash\, \aslash\, \kslash \, \bslash)}
\def\iakb{{\tr}_{+}(\islash\, \aslash\, \kslash \, \bslash)}

\def\tofo{{\tr}_{+}(\onslash\, \thslash\, \twslash \, \foslash)}
\def\foto{{\tr}_{+}(\onslash\, \thslash\, \foslash \, \twslash)}
\def\tofi{{\tr}_{+}(\onslash\, \thslash\, \twslash \, \fislash)}
\def\fito{{\tr}_{+}(\onslash\, \thslash\, \fislash \, \twslash)}

\def\lrangle#1#2{\langle #1\,#2\rangle}

\def\Li{{$\rm Li}_2$}
\def\eps{\epsilon}
\def\epsuv{{\epsilon_{\rm \mbox{\tiny UV}}}}
\let\bm=\bibitem
\let\la=\label

\def\npb#1#2#3{Nucl. Phys. {\bf{B#1}} #3 (#2)}
\def\plb#1#2#3{Phys. Lett. {\bf{#1B}} #3 (#2)}
\def\prl#1#2#3{Phys. Rev. Lett. {\bf{#1}} #3 (#2)}
\def\prd#1#2#3{Phys. Rev. {D \bf{#1}} #3 (#2)}
\def\cmp#1#2#3{Comm. Math. Phys. {\bf{#1}} #3 (#2)}
\def\cqg#1#2#3{Class. Quantum Grav. {\bf{#1}} #3 (#2)}
\def\nppsa#1#2#3{Nucl. Phys. B (Proc. Suppl.) {\bf{#1A}}#3 (#2)}
\def\ap#1#2#3{Ann. of Phys. {\bf{#1}} #3 (#2)}
\def\ijmp#1#2#3{Int. J. Mod. Phys. {\bf{A#1}} #3 (#2)}
\def\rmp#1#2#3{Rev. Mod. Phys. {\bf{#1}} #3 (#2)}
\def\mpla#1#2#3{Mod. Phys. Lett. {\bf A#1} #3 (#2)}
\def\jhep#1#2#3{J. High Energy Phys. {\bf #1} #3 (#2)}
\def\atmp#1#2#3{Adv. Theor. Math. Phys. {\bf #1} #3 (#2)}
\newcommand{\EQ}[1]{\begin{equation} #1 \end{equation}}
\newcommand{\AL}[1]{\begin{subequations}\begin{align} #1 \end{align}\end{subequations}}
\newcommand{\SP}[1]{\begin{equation}\begin{split} #1 \end{split}\end{equation}}
\newcommand{\ALAT}[2]{\begin{subequations}\begin{alignat}{#1} #2 \end{alignat}
                        \end{subequations}}
\def\beqa{\begin{eqnarray}}
\def\eeqa{\end{eqnarray}}
\def\beq{\begin{equation}}
\def\eeq{\end{equation}}
\def\sst{\scriptscriptstyle}
\def\thetabar{\bar\theta}
\def\Tr{{\rm Tr}}
\def\one{\mbox{1 \kern-.59em {\rm l}}}
 \def\Nh{\hat{N}}

\newcommand{\half}{{\textstyle {1 \over 2}}}

\def\a{\alpha}      \def\da{{\dot\alpha}}
\def\b{\beta}       \def\db{{\dot\beta}}
\def\c{\gamma}  \def\G{\Gamma}  \def\cdt{\dot\gamma}
\def\d{\delta}  \def\D{\Delta}  \def\ddt{\dot\delta}
\def\e{\epsilon}        \def\vare{\varepsilon}
\def\f{\phi}    \def\F{\Phi}    \def\vvf{\f}
\def\h{\eta}
\def\k{\kappa}
\def\l{\lambda} \def\L{\Lambda}
\def\m{\mu} \def\n{\nu}
\def\o{\omega}
\def\p{\pi} \def\P{\Pi}
\def\r{\rho}
\def\s{\sigma}  \def\S{\Sigma}
\def\t{\tau}
\def\th{\theta} \def\Th{\Theta} \def\vth{\vartheta}
\def\X{\Xeta}
\def\z{\zeta}
\def\de{\partial}

\def\cA{{\cal A}} \def\cB{{\cal B}} \def\cC{{\cal C}}
\def\cD{{\cal D}} \def\cE{{\cal E}} \def\cF{{\cal F}}
\def\cG{{\cal G}} \def\cH{{\cal H}} \def\cI{{\cal I}}
\def\cJ{{\cal J}} \def\cK{{\cal K}} \def\cL{{\cal L}}
\def\cM{{\cal M}} \def\cN{{\cal N}} \def\cO{{\cal O}}
\def\cP{{\cal P}} \def\cQ{{\cal Q}} \def\cR{{\cal R}}
\def\cS{{\cal S}} \def\cT{{\cal T}} \def\cU{{\cal U}}
\def\cV{{\cal V}} \def\cW{{\cal W}} \def\cX{{\cal X}}
\def\cY{{\cal Y}} \def\cZ{{\cal Z}}

\def\ua{\underline{\alpha}}
\def\ub{\underline{\phantom{\alpha}}\!\!\!\beta}
\def\uc{\underline{\phantom{\alpha}}\!\!\!\gamma}
\def\um{\underline{\mu}}
\def\ud{\underline\delta}
\def\ue{\underline\epsilon}
\def\una{\underline a}\def\unA{\underline A}
\def\unb{\underline b}\def\unB{\underline B}
\def\unc{\underline c}\def\unC{\underline C}
\def\und{\underline d}\def\unD{\underline D}
\def\une{\underline e}\def\unE{\underline E}
\def\unf{\underline{\phantom{e}}\!\!\!\! f}\def\unF{\underline F}
\def\unm{\underline m}\def\unM{\underline M}
\def\unn{\underline n}\def\unN{\underline N}
\def\unp{\underline{\phantom{a}}\!\!\! p}\def\unP{\underline P}
\def\unq{\underline{\phantom{a}}\!\!\! q}
\def\unQ{\underline{\phantom{A}}\!\!\!\! Q}
\def\unH{\underline{H}}

\def\As {{A \hspace{-6.4pt} \slash}\;}
\def\bs {{b \hspace{-6.4pt} \slash}\;}
\def\Ds {{D \hspace{-6.4pt} \slash}\;}
\def\ds {{\del \hspace{-6.4pt} \slash}\;}
\def\ss {{\s \hspace{-6.4pt} \slash}\;}
\def\ks {{ k \hspace{-6.4pt} \slash}\;}
\def\ps {{p \hspace{-6.4pt} \slash}\;}
\def\pas {{{p_1} \hspace{-6.4pt} \slash}\;}
\def\pbs {{{p_2} \hspace{-6.4pt} \slash}\;}
\def\Ps {{P \hspace{-6.4pt} \slash}\;}
\def\Qs {{Q \hspace{-6.4pt} \slash}\;}

\def\Fh{\hat{F}}
\def\Vh{\hat{V}}
\def\Xh{\hat{X}}
\def\ah{\hat{a}}
\def\xh{\hat{x}}
\def\yh{\hat{y}}
\def\ph{\hat{p}}
\def\xih{\hat{\xi}}
\def\psit{\tilde{\psi}}
\def\psit{\tilde{\psi}}
\def\tht{\tilde{\th}}
\def\lt{\tilde{\lambda}}
\def\hl{\hat{\lambda}}
\def\hlt{\hat{\tilde{\lambda}}}
\def\llt{\tilde{l}}
\def\At{\tilde{A}}
\def\Qt{\tilde{Q}}
\def\Rt{\tilde{R}}
\def\Nt{\tilde{N}}

\def\at{\tilde{a}}
\def\st{\tilde{s}}
\def\pt{\tilde{p}}
\def\qt{\tilde{q}}
\def\vt{\tilde{v}}
\def\nt{\tilde{n}}

\def\delb{\bar{\partial}}
\def\bz{\bar{z}}
\def\bD{\bar{D}}
\def\bB{\bar{B}}

\def\bk{{\bf k}}
\def\bl{{\bf l}}
\def\bp{{\bf p}}
\def\bq{{\bf q}}
\def\br{{\bf r}}
\def\bx{{\bf x}}
\def\by{{\bf y}}
\def\bR{{\bf R}}
\def\bV{{\bf V}}

\def\d{\delta}\def\D{\Delta}\def\ddt{\dot\delta}
\def\pa{\partial} \def\del{\partial}
\def\xx{\times}
\def\uno{\mbox{1 \kern-.59em {\rm l}}}
\def\trp{^{\top}}
\def\inv{^{-1}}
\def\dag{{^{\dagger}}}
\def\pr{^{\prime}}
\def\lan{\langle}
\def\ran{\rangle}
\def\rar{\rightarrow}
\def\lar{\leftarrow}
\def\lrar{\leftrightarrow}
\newcommand{\0}{\,\!}      
\def\one{1\!\!1\,\,}
\def\im{\imath}
\def\jm{\jmath}
\newcommand{\tr}{\mbox{tr}}
\newcommand{\slsh}[1]{/ \!\!\!\! #1}
\def\vac{|0\rangle}
\def\lvac{\langle 0|}
\def\hlf{\frac{1}{2}}
\def\ove#1{\frac{1}{#1}}
\def\Box{\square}
\def\ZZ{\mathbb{Z}}
\def\CC#1{({\bf #1})}
\def\bcomment#1{}
\def\bfhat#1{{\bf \hat{#1}}}
\def\VEV#1{\left\langle #1\right\rangle}
\newcommand{\ex}[1]{{\rm e}^{#1}} \def\ii{{\rm i}}
\def\rr{{\rm r}} \def\rs{{\rm s}}\def\rv{{\rm v}}
\def\ri{{\rm i}}\def\rj{{\rm j}}
\def\ie{{\it i.e.}}
\def\eg{{\it e.g.}}
\newcommand{\lrbrk}[1]{\left(#1\right)}
\newcommand{\sfrac}[2]{{\textstyle\frac{#1}{#2}}}

\def\Li{{\rm Li}_2}

\font\mybb=msbm10 at 12pt
\def\bb#1{\hbox{\mybb#1}}

\font\myBB=msbm10 at 18pt
\def\BB#1{\hbox{\myBB#1}}

\newcommand{\ft}[2]{{\textstyle\frac{#1}{#2}}}

\newcommand{\R}{\mathbbm{R}}
\newcommand{\Z}{\mathbbm{Z}}
\newcommand{\C}{\mathbbm{C}}

\newcommand{\dalpha}{\dot{\alpha}}
\newcommand{\dbeta}{\dot{\beta}}

\begin{document}

\begin{flushright}
ROM2F/2015/11
\end{flushright}

\vspace{8pt}

\begin{center}

{\Large \bf Instanton corrections to the effective action  }
\\
\vspace{0.4cm}
{\Large \bf    of $\mathcal{N}=4$ SYM  }

\vspace{16pt}

{\mbox {\bf  Massimo Bianchi${}$, Jose Francisco Morales$^{}$ and  Congkao Wen$^{}$}}%
\footnote{
{ \{ \tt \!\!\! massimo.bianchi, francisco.morales, congkao.wen\}@roma2.infn.it}
}


\begin{quote}
{\small \em
\begin{itemize}
\item[\ \ \ \ \ \ $^{}$]
Dipartimento di Fisica, 
Universit\`a di Roma ``Tor Vergata"  \\
\& I.N.F.N. Sezione di Roma ``Tor Vergata" \\
Via della Ricerca Scientifica, 00133 Roma, Italy
\end{itemize}
}
\end{quote}


\vspace{60pt} {\bf Abstract}
\end{center}

\noindent
We compute the one-instanton effective action of  ${\cal N}=4$ super Yang-Mills theory with gauge group $Sp({2N})$.   The result can be written in a very compact and manifestly supersymmetric form  involving an integral over the superspace of an irrational function of the ${\cal N}=4$ on-shell superfields. 
In the Coulomb branch, the instanton corrects both the MHV  and next-to-next-MHV higher derivative terms $D^4F^{2n+2}$ and $F^{2n+4}$. We confirm at the non-perturbative level  the non-renormalization theorems for MHV $F^{2n+2}$ terms that are expected to receive perturbative corrections only at $n$-loops. We compute also the one and two-loop corrections to the $D^4 F^4$ term and show that its  completion under $SL(2,\Z)$ duality is consistent with the one-instanton results of $U(2)$ gauge group.

\setcounter{page}{0}
\thispagestyle{empty}
\newpage


\setcounter{tocdepth}{4}
\hrule height 0.75pt
\tableofcontents
\vspace{0.8cm}
\hrule height 0.75pt
\vspace{1cm}

\setcounter{tocdepth}{2}


\setcounter{footnote}{0}

\section{Introduction  }
\label{intromotiv}

Instanton effects play an intriguing role in Yang-Mills theory. Starting from the seminal work by Belavin, Polyakov, Schwartz and Tyupkin \cite{Belavin:1975fg} and by 't Hooft \cite{'tHooft:1976fv}, the understanding of instanton effects has deepened and deepened, especially in supersymmetric theories\footnote{See e.g. \cite{Bianchi:2007ft} for a recent review.}. In particular in ${\cal N} =1$ supersymmetric theories, instantons are known to generate non-perturbative super-potentials, thus violating perturbative non-renormalization theorems and triggering a host of interesting effects, including dynamical supersymmetry breaking. In ${\cal N} =2$ theories, they correct the analytic pre-potential that can be computed `exactly' using localisation techniques. In ${\cal N} =4$ supersymmetric Yang-Mills theories (SYM) as well as in other exactly super-conformal theories, the situation is trickier. Essentially nothing was known before Maldacena's conjecture \cite{Maldacena:1997re}. With the advent of AdS/CFT, the correspondence between Yang-Mills instantons and D-instantons has been established supporting the validity of the holographic conjecture beyond the perturbative regime \cite{Bianchi:1998nk}. Instantons correct correlation functions of scaling operators, in particular of chiral primary operators (CPO's). As a consequence, they contribute to the anomalous dimensions of certain double- and multi-trace operators \cite{Bianchi:1999ge}. No instanton corrections to the cusp anomalous dimension are expected, neither to Konishi-like operators \cite{Bianchi:2001cm, Bianchi:2000hn}. As shown in \cite{Bianchi:2013xsa}, instanton corrections to correlation functions of CPO's in ${\cal N} =4$ SYM vanish in the pairwise light-like limit, which hints to the absence of such corrections to light-like Wilson loops if the so-called correlation/Wilson loop duality~\cite{Alday:2007hr, Drummond:2007aua, Brandhuber:2007yx, Alday:2010zy, Eden:2010zz} still holds at the non-perturbative level. Moreover, just on dimensional grounds, higher-derivative terms cannot get corrected by instantons in the super-conformal phase since there is no dimensionful constant in this phase to build such terms.

It is thus mandatory to ask whether instantons correct the effective action of  ${\cal N} =4$ SYM and scattering amplitudes/ higher derivative terms in the non-conformal phases of the theory. In this paper we show that the answer to both questions is positive. 
 We will  study corrections to the effective action of  ${\cal N} =4$ SYM and to higher derivative terms in the Coulomb branch of the theory.  There has been recent attention on the study of  ${\cal N}=4$ SYM theory in the Coulomb branch. 
 The motivations for this study are two-fold. On the one hand, turning on a vacuum expectation value (vev) for the scalar fields in ${\cal N}=4$ SYM theory provides an IR regularisation compatible with Poincar\`e supersymmetry and preserving the dual-conformal symmetry of the theory in the the planar limit \cite{Alday:2009zm}. On the other hand, the theory on the Coulomb branch can be studied in its probe approximation via AdS/CFT tools. This observation motivated the recent proposal by Schwarz \cite{Schwarz:2013wra} so-called ``Highly Effective Action" (HEA) of ${\cal N}=4$ SYM with gauge group $U(2)$   {}\footnote{ For earlier proposals on the effective action of ${\cal N}=4$ SYM theory, see for instance  \cite{GonzalezRey:1998uh, Buchbinder:1999jn, Kuzenko:2000tg}.}.
  
The HEA was derived from the Dirac-Born Infeld action of a single brane viewed as a probe of the $AdS_5 \times S^5$ near horizon geometry generated by a second one. The resulting action was shown to be invariant under $SL(2,\Z)$ strong-weak coupling duality of the ${\cal N}=4$ SYM.  Still, this formulation of 
the HEA cannot be complete. It was already pointed out in ~\cite{Buchbinder:2001ui} that one-loop results of general non-MHV operators $F_+^n F_-^m$ with $m,n$ both equal or larger than four are different from those predicted by the HEA~\footnote{Here $F_{\pm}$ denote the self-dual and anti-self-dual field strengths, and we borrow the terminology from scattering amplitudes by calling operators with two $F_-$ as maximally helicity violating, namely MHV in short.}. Moreover the HEA does neither include terms involving derivatives of $F$ (and their SUSY related terms), nor instanton corrections. The fact that the DBI probe action on $AdS_5 \times S^5$ background does not capture the contributions of instantons is not surprising since instantons correspond to the inclusion of D(-1)-branes that certainly back-react on the AdS geometry. 
    
   In this paper we derive the effective action of the ${\cal N}=4$ theory with gauge group  $Sp(2N)$ at the one-instanton level.  The choice of the gauge group is motivated by the fact that the single instanton for an $Sp(2N)$ gauge theory is described by a moduli space with $O(1)$ group 
   structure and no ADHM constraints.  We  derive a manifestly supersymmetric effective action generated by a single instanton in the background of on-shell ${\cal N}=4$ superfields. The effective action codify one-instanton corrections to a large class of higher-dimensional terms in the Coulomb branch of the theory.  The emerging effective action bears some resemblance with the one found  by Green and Gutperle in \cite{Green:2000ke}, when considering  higher derivative D-instanton corrections to the ${\cal N}=4$ DBI theory for a single D3-brane. In our case the role of $\alpha'$ is played by the vev of the scalar fields.  
   
  It is interesting to ask how our results are compatible with the expected $SL(2,\Z)$-invariance of the quantum theory. 
  This symmetry, combined with explicit perturbative results, completely fixes the form of similar higher-derivative terms in the $U(1)$ theory \cite{Green:2000ke} to be given by modular forms of specific weights  that sum up a full tower of instanton corrections. In the present case the situation is more involved. Unlike the cases in \cite{Green:2000ke}, the higher derivative terms under consideration receive corrections at different loop orders and  their $SL(2,\Z)$ completion is less obvious. For the simplest MHV coupling $D^4F^4$ we derive the one-loop and two-loop corrections as well as their $SL(2,\Z)$ completion obtained by summing over their $SL(2,\Z)$ images for U(N) gauge group. The resulting  $SL(2,\Z)$  invariant function includes one-instanton contributions compatible with those we find as well as an infinite tower of instanton corrections. 
  Still, to determine the exact form of the coupling an explicit evaluation of its higher-loop  corrections is needed. 
  
   As a byproduct of our analysis, we show that the MHV terms $F_-^2 \,F_+^{2n}$ do not receive one-instanton corrections, which confirms that the non-renormalization theorems proven in~\cite{Chen:2015hpa} do hold at the non-perturbative level. Due to tight constraints from $\mathcal{N}=4$ supersymmetry, these terms are special
   since they receive perturbative corrections only at $n$-loops, and the coefficients of these terms match precisely with those coming from the HEA~\cite{Chen:2015hpa}.

 The plan of the paper is as follows. In section \ref{section:N=4superfields}, we will briefly review the ${\cal N}=4$ on-shell super fields. In Section \ref{section:One-instanton corrections}, we exploit the unoriented open string construction to derive  the exact one-instanton generated effective
 action of ${\cal N}=4$ theory with gauge group $Sp({2N})$ or $U(2)$.  In Section \ref{section:Coulomb branch},  expanding the action  in the Coulomb branch, we study  MHV and non-MHV higher derivative terms in the supersymmetric class of $D^4 F^4$ and $F^{2n}$ respectively.  
  In Section \ref{section:D^nF^4}, we compute perturbation contributions to the coupling $D^4 F^4$ (more generally $D^n F^4$) at one and two loops.
  In Section \ref{section:SL(2,Z)}, we  write the $SL(2,\Z)$ completion of the one and two-loop results for $D^4 F^4$, and show that the result is compatible with those following from the one-instanton effective action.  Finally, in Section \ref{section:conclusion}, we summarize our results and comment on possible future research directions.

\section{The ${\cal N}=4$  superfields } \label{section:N=4superfields}

In this section we review the on-shell superfield formulation of $\cN=4$ SYM theory. 
 The field content of $\cN = 4$ theory includes a vector $A_{\alpha\dot\alpha}$, four gaugini $(\psi_\alpha^A,\bar\psi_{\dot\alpha A})$ and six scalar fields 
  $\varphi_{AB}$. Indices $\alpha,\dot\alpha=1,2$  label the two chiral spinor representations  of the Lorentz group $SO(4)\sim SU(2)_{\rm L} \times  SU(2)_{\rm R} $ while $A=1,\ldots, 4$ runs over the spinor representations of the R-symmetry group $SO(6)\sim SU(4)$.  The $4\times4$ matrix  $\varphi_{AB}$ is antisymmetric and satisfies the reality
  condition
\bea
\varphi^{AB}=\varphi_{A B}^\dagger = {1\over 2} \varepsilon^{ABCD} \varphi_{CD}  
\eea
leaving six real degrees of freedom. All fields transform in the adjoint of the gauge group and fill a single (ultra-short) multiplet of the ${\cal N}=4$ supersymmetry algebra.

  It is convenient to package the elementary fields  into superfields.  It is well known that no off-shell super field description of $\cN = 4$ SYM is available that manifestly preserves full $\cN =4$ supersymmetry with a finite number of auxiliary fields.  For our purposes, \ie ~ for computing corrections to the on-shell effective action, the on-shell $\cN=4$ superspace is suitable. 
 
The basic superfield is defined by~\cite{Green:2000ke}
\bea
 {\cal W}_{A B}(x, \theta,\bar \theta) &=& 
\varphi_{AB}(x)+ \bar \psi_{[A} \, \bar\theta_{B]} +  {1 \over 2}\epsilon_{ABCD} \theta^{C} \, \psi^{D} -\bar\theta_{[A} \, F^+\, \bar\theta_{B]}-{1 \over 2} \epsilon_{ABCD} \theta^{C} F^- \, \theta^{D}   
 +\ldots \nonumber
\eea
with $\theta^A_\alpha$, $\bar\theta_{A\dot\alpha}$ the Grassmann coordinantes, and $F^-_{\alpha\beta}=\ft12\sigma^{mn}_{\alpha\beta} F_{mn}$, $F^+_{\dot\alpha\dot\beta}=\ft12 \bar\sigma^{mn}_{\dot\alpha\dot\beta} F_{mn}$ the 
antiself- and self-dual components of the field strength respectively. 
Other superfields can be constructed acting with super-derivatives on ${\cal W}_{A B}$ 
\be
 {\cal W}_{  \dot{\alpha} A} 
=  \bar{D}^{B}_{  \dot{\alpha}} {\cal W}_{AB} \, ,  \quad\quad~~~~~~~~~~~~~~
 {\cal W}_{\dot\alpha\dot\beta}  = \ft12
\bar{D}^{A}_{\dot{\alpha}} \bar{D}^{B}_{\dot{\beta}} {\cal W}_{AB} \, ,
 \ee
and so on.  Alternatively one can use the
twistor coordinates $(\lambda_\alpha ,\tilde \lambda_{\dot\alpha},\eta_A)$ of the supersymmetry algebra $SU(2,2|4)$ and write  the on-shell superfield components as (in momentum space)
\bea
 F^+_{\dot \alpha\dot\beta}& =& \tilde \lambda_{\dot\alpha}   \, \tilde \lambda_{\dot\beta}   \qquad~~~~~~  \bar\psi_{\dot\alpha A}=\tilde \lambda_{\dot\alpha}   \,\eta_A \qquad~~~~~ 
 \varphi_{AB}=\eta_A \,\eta_B   \nn\\
\psi_\alpha^A &=& \ft{1}{3!}\epsilon^{ABCD}\eta_B \,\eta_C \, \eta_D\, \lambda_\alpha \qquad ~~~~~~~~~~~~ 
 F^-_{ \alpha\beta} =\ft{1}{4!}  \,\epsilon^{ABCD}\, \eta_A\, \eta_B \,\eta_C \, \eta_D    \, \lambda_\alpha \, \lambda_\beta
 \eea
 so that the on-shell superfields read
 \bea
 {\cal W}_{A B} (x, \theta,\bar \theta) &=&   e^{ \theta  \, \lambda  \, \eta + \bar\theta  \, \tilde \lambda {\partial \over \partial \eta}  } \, \eta_A \,\eta_B \nn\\
  {\cal W}_{  \dot{\alpha} A}(x, \theta,\bar \theta)  &=&  e^{ \theta  \, \lambda  \, \eta + \bar\theta  \, \tilde \lambda {\partial \over \partial \eta}  } \,   \,\tilde \lambda_{\dot\alpha}   \,\eta_A      \nn\\
 {\cal W}_{\dot\alpha\dot\beta} (x, \theta,\bar \theta) &=&     e^{ \theta  \, \lambda  \, \eta + \bar\theta  \, \tilde \lambda {\partial \over \partial \eta}  } \,   \tilde \lambda_{\dot\alpha}   \,\tilde \lambda_{\dot\beta} 
\eea
Henceforth we use the standard spinor-helicity notation for momentum, and Lorentz invariants~\footnote{See for instance \cite{Elvang:2013cua} for a recent review on the spinor-helicity formalism and related topics.}
 \bea
p^{\alpha \dot\alpha}_i&=& \lambda^\alpha_i \,\tilde \lambda^{\dot\alpha}_i  \,,      \qquad \langle i\, j \rangle = \epsilon_{\alpha\, \beta }\lambda^{\alpha}_i\,  \lambda^{\beta}_j  \, ,   
\qquad \left[  i\, j \right]  =\epsilon^{\dot\alpha\, \dot\beta }  
(\tilde{\lambda}_i)_{\dot\alpha}\,  (\tilde{\lambda}_j)_{\dot\beta}   \, .
 \eea

Finally we recall that the classical action of the $\cN=4$ theory is invariant under S-duality transformations $S: \tau \to -{1\over \tau}$ (here $\tau  = {\vartheta \over  2\pi}+\ii {4\pi \over g }$) acting on the field theory variables as
\bea
 F^+ & \to &  e^{i \alpha \over 2}  \, F^+    ~,~   \bar\psi \to   e^{i \alpha \over 4}\, \bar\psi   ~,~  
 \varphi \to     \varphi  ~,~ \psi \to    e^{- {i \alpha \over 4} }\, \,  \psi    ~,~
F^-  \to    e^{- {i \alpha \over 2} } \, F^-     \label{fdual}
   \eea
 with   
  \be
  e^{i \alpha } \equiv {\tau \over \bar\tau} \, .
  \ee 
  Together with $\tau\to \tau+1$, this generates an $SL(2,\Z)$ group that is expected to be an exact symmetry of the full quantum theory. On the twistor coordinates S-duality acts according to
 \be
S:  \quad\quad  \tilde \lambda \to   e^{ {i \alpha \over 4} } \,  \tilde \lambda  \, ,  \qquad ~~~~   \lambda\to  e^{- {i \alpha \over 4} }   \,  \lambda  \, , \qquad~~~ \eta  \to    \eta \, .  \label{sl2zu}
 \ee

\section{Effective Theory: One-instanton corrections}   \label{section:One-instanton corrections}

The ${\cal N}=4$ SYM in 4D can be realised in String Theory as the low energy theory describing the dynamics of open strings ending on a stack of 
D3-branes. For $N$ D3-branes and oriented open strings one finds the gauge group $U(N)$. Instantons can be realised by the inclusion of D(-1)-branes. 
 The lowest modes of open strings with at least one end on the D(-1)-branes exactly produce the moduli (positions, size, orientations)  specifying the instanton solution  and the D(-1)-D3 action describes the Yang-Mill action in the instanton background. 
Among many things, there are a few main advantages of working with open strings in the D(-1)-D3 system which we will remark shortly. 
First, one has an explicit prametrization of the instanton super-moduli space. Second, by computing string amplitudes on disks with mixed boundary conditions (both D3 and D(-1)), one can derive the exact couplings of the super-moduli to the physical on-shell fields. Moreover, including unoriented projections that combine world-sheet parity $\Omega$ with space-time involutions one can analyse orthogonal and symplectic groups. Last but not least, orbifold projections  allow to break superymmetry to $\cN=2$ or $\cN=1$ as well as the gauge group and get interesting  quiver (conformal or non-conformal) gauge theories.

\subsection{The instanton action}

Using the D(-1)-D3 system, instanton moduli are organised  according to their transformation properties with respect to the symmetry group $U(k)\times U(N)\times SU(2)^2\times SU(4)$. Here $U(k)$ is the instanton symmetry of $k$ D(-1)'s, $U(N)$ is the gauge symmetry of $N$ D3-brane's, $SU(2)^2\sim SO(4)$ is the Lorentz group (in the directions longitudinal to the D3's), and finally $SU(4)\sim SO(6)$ is the R-symmetry group (acting on the directions transverse to the D3's). The moduli associated to massless modes of D(-1)D(-1) and D(-1)D3 open strings are summarized as follows, 
  \bea
{\rm D(-1)D(-1)} : && {\bf V}_{k}= \{  \chi_{AB},   D_{\dot \alpha}^{\dot \beta} ;  \tilde\Theta_{\dot\alpha A}   \}_i^j  \quad\quad 
~~~~~~~~~~~({\bf \bar k}, {\bf k}) \nonumber\\ 
&& {{\bf H}_{\rm adj}}_k =\{ a_{\alpha\dot\alpha} ; \Theta_{\alpha}^A    \}_i^j  \quad\quad  ~~~~~~~~~~~~~~~~({\bf \bar k}, {\bf k})   \nonumber\\
{\rm  D(-1)D3} :&& {\bf H}_{\rm bif}=\{ w_{\dot\alpha} , \mu^A \}^{i}_u ~, \{ \bar w_{\dot\alpha} , \bar\mu^A \}_{i}^u    ~~~~~~~~~({\bf \bar k}, {\bf N})+({\bf \bar N}, {\bf  k}) \,. \label{moduli}
\eea
The indices $ i=1, \ldots ,k$, $u=1,\ldots N$, $\alpha,\dot\alpha=1,2$  and $A=1, \ldots, 4$ run over the fundamental representations 
of  the various symmetry groups. Upper and lower indices refer to the fundamental and anti-fundamental representations respectively for groups
different from $SU(2)$ whose indices are raised and lowered with the $\varepsilon$-tensor.  The fields $a_{\alpha\dot\alpha}$ and $\chi_{[AB]}$ parametrise the positions of the instantons along the directions longitudinal and transverse to the D3-branes respectively. The field 
$D_{\dot \alpha}^{\dot \beta}=-\ii D^c (\tau^c)_{\dot \alpha}^{\dot\beta}  $ with $(\tau^c)_{\dot \alpha}^{\dot\beta} $ the Pauli matrices, 
  is an auxiliary field acting as Lagrangian multiplier for
the $3k^2$ ADHM constraints, while $w_{\dot\alpha}$ and $\bar w_{\dot\alpha}$ describe the massless modes of the D(-1)D3 strings. Finally the fermionic fields $\Theta_{\alpha}^A,  \tilde\Theta_{\dot\alpha A} $ as well as $\mu^A,\bar \mu^A$ are supersymmetry partners. 

 The fields in (\ref{moduli}) have been conveniently grouped into vector ${\bf V}$ and hypermultiplet ${\bf H}$ representations of ${\cal N}=(1,0)$ supersymmetry 
   in six dimensions. The  instanton action  is given by the dimensional reduction of the six-dimensional $U(k)$ gauge theory minimally coupled to the hypermultiplets \cite{Dorey:2002ik}. We notice that the center of mass $(a,\Theta)$ of the adjoint hypermultiplet always decouples from the rest of the theory. They represent   the super-coordinates of the instanton. In the absence of a background for D3-D3 fields the instanton partition trivially vanishes since the 
   action is independent of $\Theta$.  To find a non-trivial result we consider instead the instanton partition function in the background of the 
   ${\cal N}=4$ superfields 
     \bea
{\rm D3D3} : && {\bf V}_{N}= \{  \Phi_{AB};  \Lambda_{\dot\alpha A} ;   {\cal F}_{\dot \alpha}^{\dot \beta}    \}_u^v  \quad\quad 
~~~~~~~({\bf \bar N}, {\bf N})  \label{backg}
\eea
with
    \be
\Phi_{AB} ={\cal W}_{AB}(a,\Theta,0)  \quad\quad \Lambda_{\dot\alpha A}  ={\cal W}_{\dot\alpha A}(a,\Theta,0) 
 \quad\quad   {\cal F}_{\dot \alpha}^{ \dot \beta}  ={\cal W}_{\dot \alpha}^{\dot \beta}(a,\Theta,0)   \label{phiw}
\ee
the basic ${\cal N}=4$ superfields  (\ref{phiw}) evaluated at the instanton super-coordinates. We notice that D3-D3 superfields in (\ref{backg}) fill the components of 
a $U(N)$ vector multiplet of the  ${\cal N}=(1,0)$ supersymmetry in six dimensions, so the instanton action now follows from a reduction of  a 
$U(k)\times U(N)$ gauge theory.  The instanton action can then be read from formulae (10.70bc) in \cite{Dorey:2002ik} after the 
replacement  $\chi\to  g( \chi+\Phi) $, $\tilde \Theta \to g (\tilde \Theta +\Lambda)$ and $D\to g( D+{\cal F}) $. The shifts account for the
extra couplings to ${\cal N}=4$ superfields while the extra powers of $g$ are introduced here to match the field theory conventions we will later use 
in the computation of loop corrections to higher derivative terms\footnote{In these conventions
  the Yang-Mills lagrangian reads ${\cal L}=-\ft14 F^2+\ldots$ and covariant derivatives are defined as $\nabla=\partial +\ii g \, A$. }.
 For $k=1$ one  finds
\bea
&& S_{\rm inst} = {4\pi^2\over g^2} \left[ g^2\,  \bar w^{u\dot\alpha} w_{v \dot\alpha} 
(  \chi_{AB}\,\delta_u^{v'} + \Phi_{ABu}^{v'} ) (  \chi_{AB}\,\delta_{v'}^{v} + \Phi_{AB v'}^{v} )
+\sqrt{2}  \, g\,\bar \mu^{Au} \mu^B_v\, \left(  \chi_{AB} \, \delta_u^v+ \Phi_{ABu}^v  
 \right)   \right.  \nn\\
&& \left. + \, g\, \bar w^{u\dot\alpha} w_v^{\dot\beta} \left( D_{\dot \alpha}^{ \dot \beta}\, \delta_u^v+ {\cal F}_{\dot \alpha u}^{\dot \beta v}  
 \right) 
 +  \ii  \,g\,\left(\bar{\mu}^{Au} w_{\dot{\alpha}v}
+\bar{w}^u_{\dot{\alpha}}\mu_v^{A}\right) \left( \tilde\Theta^{\dot{\alpha}}_{A}\, \delta_u^v+ \bar\Lambda^{\dot{\alpha}v}_{Au}  
 \right)   \right] \, . \label{sinst} 
\eea
 We notice that the $k=1$ instanton action (\ref{sinst}) is bilinear in the D(-1)-D3 fields since only these couple non-trivially to the $U(1)$ theory. 
  Alternatively, the action (\ref{sinst}) can be derived by computing string amplitudes on disks with mixed boundary conditions \cite{Billo:2002hm,Billo:2008pg}. The cubic moduli interactions are given by $\langle V_\chi V_{\bar\mu} V_{\mu} \rangle$, $\langle V_D V_{\bar w} V_w  \rangle$ and $\langle V_{\tilde \Theta} V_\mu V_{\bar\mu} \rangle$ with the vertex operators given by
\bea
V_\chi &=&\chi_{AB}\, e^{-\varphi} \,  \psi^{AB} \, ,   \quad\quad~~~~~~~    V_{D}= D_{\dot \alpha \dot\beta}\,  \sigma_{\mu\nu}^{\dot\alpha\dot\beta} \, \psi^\mu \psi^\nu \, ,
 \quad\quad~~~~~~~  
  V_w  =  w^{\dot \alpha} \, e^{-\varphi}   \, C_{\dot\alpha} {\Sigma} \, ,
  \cr
  V_\mu&=&  \nu^A e^{-\varphi/2} \, C_{A} \, {\Sigma} \, , \quad\quad ~~~~~~\, V_{\bar\Theta}  = \tilde \Theta^{\dot \alpha}_A \, e^{-\varphi/2}\, C_{\dot\alpha} S^A    \,, 
\eea
where $C,S$ denote various spin fields of opposite chirality and ${\Sigma}$ is a twist-field in the four Neumann-Dirichlet directions. Complex conjugate fields are given by the same vertex operators with opposite orientation of their Chan-Paton matrices. The same three-point functions determine the couplings to the lowest components of the D3-D3 superfields, with the same string vertices $V_\chi$, $V_D$ and $V_{\bar\Theta}$ but
 now inserted on the D3-boundary. Finally, quartic interactions  $\Phi^2\, w\, \bar w$ can be derived from the three-point function 
 $\langle V_{X} \, V_\chi\,V_w \rangle$, with $V_X=   X_{AB}^{\dot \alpha} \,  \psi^{AB}   \, C_{\dot\alpha} {\Sigma} $ an auxiliary field  with standard kinetic term 
 $X \bar X$. Solving for $X$, the quartic interaction is reproduced. We remark also that the vertices $V_D$ and $V_X$ are not BRST invariant as expected for auxiliary fields.  The $D$ field implements the  $3$ ADHM constraints for the case of one instanton.   

  The effective action generated by the D-instanton  is defined by the integral over the instanton moduli space \cite{Dorey:2002ik}
\be
S^{k=1}_{\rm eff}= c\, {g^{4N + 4}\over \pi^{6N+6} } \,e^{2\pi \ii \tau }\, \int d^4a\, d^8\Theta\,  d^{6} \chi \, {d^3 D \over (2 \pi)^3} \, d^8\tilde \Theta\, d^{2N} w\, d^{2N} \bar w \, d^{4N}\mu \, d^{4N}\bar \mu\, e^{-S_{\rm inst} } \, ,
\ee
where we made explicit the dependence on $\pi$ and $g$ of the instanton measure and denote by $c$ the remaining numeric factor. The integration over Lagrangian multipliers $D$ and  $\tilde \Theta$ leads to standard ADHM constraints. More precisely  we wrote the Yang-Mills integral
as
 \be
\int_{k=1} dA \,d\Psi e^{-S(A,\Psi)}     = c \,  
 {g^{n_F-n_B}\over \pi^{n_F-{1\over 2} n_B} }\,  e^{2\pi \ii \tau } \, \int  d\mathfrak M_{k=1}   \,   e^{-S_{\rm inst}}   \label{measure}
\ee
  with $n_F=8N + 8$, $n_B=4N + 4$ the numbers of bosonic and fermionic moduli. The contribution $(g /\pi)^{n_F-n_B}$ comes
  from the Jacobian that relates the Yang-Mills field differential to the moduli space volume form  (see for example \cite{Vandoren:2008xg}) and the
  extra $(2\pi)^{-n_B/2}$ from the measure of the bosonic moduli~\cite{Bianchi:2007ft}.

\subsection{ One-instanton in the $Sp({2N})$ gauge theory}

It is well known that in addition to D3-branes, preserving half of the super symmetries of the Type II theories, one can include 1/2 BPS orientifold planes $\Omega$3. The resulting theory is a close relative to Type I theory.  Open strings are unoriented and the resulting gauge groups include orthogonal or symplectic factors as well as symmetric or anti-symmetric irreps of unitary gauge groups\footnote{For D3-branes, there exist 4 different $\Omega$3-planes depending on the quantised values of $B_2$ and its R-R dual $C_2$. In addition to the standard $\Omega 3^-$ with $B_2=C_2 = 0$ producing $SO(2N)$ there exist  $\Omega 3^+$ with $B_2= \ft12$, $C_2 = 0$ producing $Sp(2N)$, $\tilde\Omega 3^-$ with $B_2=0$, $C_2 =\ft12$ producing $SO(2N+1)$ and finally $\tilde\Omega 3^+$ with $B_2=C_2 =\ft12$ producing $Sp(2N)$ (with a different dyonic spectrum wrt to $\Omega 3^+$) \cite{Bianchi:1991eu, Bianchi:1997rf, Witten:1997bs, Dudas:2001wd, Bachas:2008jv}.}.

The un-oriented projections produce `complementary' gauge groups on the D(-1)'s, orthogonal
$O(k)$ symmetry groups for $Sp({2N})$ gauge theories 
and symplectic $Sp(K)$ groups (K even) for $SO(N)$ theories. 
      For a single D-instanton, the $Sp({2N})$ case is simpler since the instanton symmetry is $O(1)$, so no ADHM constraint survives the unoriented projection.
  In this section we compute the effective action induced by the a single D-instanton  on the $Sp(2N)$ gauge theory.   
  
   The instanton moduli space for the $Sp(2N)$ gauge theory can be found by projecting (\ref{moduli}) onto $\Omega$-invariant states.
   Besides the flip of the orientation of the strings, the $\Omega$3-projection acts by a $\Z_2 \in SU(2)_\alpha$ reflection, so fields 
   with an index $\alpha$ are projected onto symmetric (singlet) and without an index $\alpha$ into anti-symmetric (empty) representations  of the $O(1)$ instanton group.   As a result  the fields  $(a_{\alpha\dot\alpha},\Theta_\alpha^A)$ are projected onto the singlet, while 
    $(\bar\Theta_{\dot \alpha A},\chi_{AB},D_{\dot\alpha\dot\beta})$  are projected out.   The action of $\Omega$ on the other ADHM data identifies D3-D(-1) strings with D(-1)-D3 strings, viz.
\be
\bar{w}^{u}_{\dot\alpha} = {w}^{u}_{\dot\alpha} \quad , \quad \bar{\mu}^{u A} = {\mu}^{u A}
\ee
The position of the indices $u$ is irrelevant since they can be raised and lowered by  a symplectic `metric' tensor.  The   one-instanton effective action becomes
\bea  \label{effect1}
S^{k=1}_{\rm eff}  = c  \,  {g^{2N+4} \over \pi^{3N+6} } \, e^{2\pi \ii \tau } \,\int d^4 a \,d^8 \Theta\,  d^{4N} \mu \,d^{2N} w \, e^{-  \,S_{inst}} \, , 
\eea
 where we made use of (\ref{measure}) with $n_F=4N+8$, $n_B=2N+4$. The instanton action reduces to
\be
S_{\rm inst} ={4\pi^2\over g^2} \left[  g^2 \, w^{\dot\alpha}  \Phi_{AB} \Phi^{AB} \,   w_{\dot\alpha} 
+\sqrt{2} \, g\,\mu^{A} \Phi_{AB} \mu^B +   g\, w^{\dot\alpha}  {\cal F}_{\dot \alpha}^{ \dot \beta }  w_{\dot\beta} 
+ 2  \ii\, g\, \, \mu^A\, \bar\Lambda^{\dot{\alpha}}_{A}    w_{\dot{\alpha}} \right] \, .   \label{sinst2}
 \ee
 The integral over $\mu^A_u$ leads to the Pfaffian $ \sqrt{{\rm det} (\Phi)_{Au,Bv} }$,   
 and produces the new term  $ \ft{4\pi^2}{g\sqrt{2} } w  \, \bar{\Lambda} \Phi^{-1} \bar{\Lambda}\,  w$ in the action. The integral over $w$'s is gaussian and contributes a square root determinant  in the denominator. The result reads
\bea   
S^{k=1}_{\rm eff} =  c' \, {g^4\over \pi^6} \, e^{2\pi \ii \tau } \, \int  { d^4 a\, d^8 \Theta\, \sqrt{{\rm det}_{4N} \, 2\Phi_{Au,Bv} }  \over 
\sqrt{{\rm det}_{2N} \left( \Phi^{AB}\Phi_{AB}  +{1\over g} {\mathcal{F} }  +\ft{1}{\sqrt{2} g} \bar{\Lambda}_{A} (\Phi^{-1})^{AB} \bar{\Lambda}_{B}  \right)_{\dot\alpha u, \dot\beta v} }  } \, . 
\label{seff0}
\eea
We cannot refrain from immediately noticing that the integrand is dimensionless as expected for an exactly conformal field theory. 
 This remarkable simple formula describes all MHV and non-MHV terms generated at one-instanton level in the ${\cal N}=4$ effective action  involving fields contained in the ${\cal N}=4$
on-shell  superfields 
{\small
\bea
 \Phi_{A B}(x, \Theta) &=& 
\varphi_{AB}(x)+{1 \over 2}\epsilon_{ABCD} (\Theta^{C} \, \psi^{D}) 
-{1 \over 2} \epsilon_{ABCD} (\Theta^{C} \, F^-\, \Theta^{D} ) 
\label{superfields}\\
 \bar\Lambda_{  \dot{\alpha} A}(x, \Theta)
&=&   \bar{\psi}_{\dot\alpha  A}
- 4i\, \Theta^{B \alpha}  {\nabla}_{\alpha \dot\alpha} \varphi_{AB} - 
2 i \epsilon_{ABCD}  \Theta^{B \alpha}   (\Theta^C {\nabla}_{\alpha\dot\alpha} \psi^D) + 
i \epsilon_{ABCD}  \Theta^{B\alpha} \Theta^{C\beta}   \Theta^{D\gamma}
\,{\nabla}_{\alpha\dot\alpha} F^{-}_{\beta\gamma} \cr
 {\cal F}_{\dot\alpha\dot\beta} (x, \Theta)&=&  
   F^+_{\dot\alpha\dot\beta}  + i \Theta^{A\alpha} {\nabla}_{\alpha\dot\alpha} \bar{\psi }_{\dot\beta A}
+ 4 \Theta^{A\alpha} \Theta^{B \beta}  {\nabla}_{\alpha\dot \alpha} {\nabla}_{\beta\dot\beta} \varphi_{AB}
\cr
&& +
2 \epsilon_{ABCD} \theta^{A\alpha} \Theta^{B\beta} \Theta^{C\gamma} {\nabla}_{\alpha\dot\alpha} {\nabla}_{\beta\dot\beta} \psi_{\gamma} +
\epsilon_{ABCD}  \Theta^{A\alpha} \Theta^{B\beta} \Theta^{C\gamma}   \Theta^{D \sigma}
{\nabla}_{\alpha\dot\alpha} {\nabla}_{\beta\dot\beta} F^-_{\gamma \sigma}   \nonumber
\eea
}
Note that the $\Theta$-expansions stops at second, third and fourth orders respectively  because of the field equations. 
     The same strategy can be applied to the study of higher derivative terms in the ${\cal N}=4$ effective action. To this aim one should
    first compute  the higher derivative couplings of D3-D3 fields to the D-instanton. Disk amplitudes computing higher derivative couplings 
    of closed string fields to the D-instanton action have been recently computed in \cite{Billo:2012st} in the context of holography.

 \subsection{ One-instanton in the $U(2)$ gauge theory}

 In the case of $SU(2)$ gauge theory we have three closely related choices: $SU(2)$, $Sp(2)$ and $SO(3)$. The three algebras
  coincide and also the theories at the perturbative level but the spectrum of dyoinc states is different.  The $SU(2)$ case is realised in terms of oriented strings while $Sp(2)$ and $SO(3)$ requires an $\Omega$-projection.  It is instructive to rederive the one-instanton effective action
 for   $SU(2)$ using the ADHM description of unitary (rather than symplectic) gauge groups.  The main difference with the $Sp(2)$ case is that now the fields $(\chi,D,\tilde\Theta)$ are part of the moduli space and the integration over $D$ and $\tilde \Theta$ produce the super ADHM constraints  
\bea
   \bar w_{\dot\alpha}{}^{u} \, (\tau^c)_{\dot\beta}^{\dot\alpha}\,    w_{u}{}^{\dot\beta} &=& 0 \, ,
   \cr
  \bar{\mu}^{Au}\, w_{u \dot{\alpha} }
+\bar{w}_{\dot{\alpha} }{}^u\,\mu_u^{A} &=& 0 \, .  
\eea
   The two equations can be solved by taking
     \bea
    w_u{}^{\dot\alpha}  &=& \rho \, {\cal U}_{u}{}^{\dot\alpha}\, ,   
    \quad\quad~~~~~~~~\bar w_{\dot\alpha}{}^u  = \rho \, {\cal U}_{\dot\alpha}{}^u \, ,
    \quad\quad~ {\cal U} \in SU(2)\,, \nn\\
     \mu^A_u &=& w_u{}^{\dot\alpha}  \, \xi^A_{\dot\alpha} \, , \quad\quad~~~~~~  \bar\mu^{Au} = \xi^{A\dot\alpha}\, \bar{w}_{\dot\alpha}{}^u \, .
  \eea
   or equivalently   $w_u{}^{\dot\alpha}=\bar   w_u{}^{\dot\alpha}$,  $\mu^A_u=\bar  \mu^A_u$. 
    Moreover the $U(1)$ D-instanton gauge symmetry  can be fixed by taking $\rho$ to be real. 
   The integrals over $D$, $\tilde{\Theta}$, $\chi$, $\bar w$ and $\bar \mu$ lead then to the constant
   \be
   {g^{12}\over \pi^{18} } \,d^6 \chi \,d^4 \bar w \,d^{8} \bar \mu  \, { \delta^3 (  {4\pi^2\over g}  \bar w \, (\tau^c)   w    ) \over {\rm vol}~U(1) } \,
 \delta^{(8)}\left(  {4\pi^2\over g}   ( \bar{\mu}^{A}\, w_{\dot{\alpha} }
+\bar{w}_{\dot{\alpha} } \,\mu^{A})  \right) e^{-4\pi^2\,  \chi^2 \rho^2} \sim   {g^{8} \over \pi^{12}} \,, 
   \ee 
   with ${\rm vol}~U(1)={2\pi/ g }$.  Thus after preforming the above integration, the $U(2)$ one-instanton effective action 
   precisely leads to (\ref{effect1}) with $N=2$.

 \section{  Instanton corrections on the Coulomb branch }   \label{section:Coulomb branch}
 
 In this section we consider the ${\cal N}=4$ effective action in the Coulomb branch.  
  The Coulomb branch is defined by taking 
\be
\langle \varphi_{AB} \rangle= v_{AB} \quad\quad ~~~~~~~~~   F^{\pm}=\lambda=\bar \lambda=0 \label{coulomb}
\ee
with $v_{AB}$ a constant matrix belonging to the Cartan subalgebra of the gauge group. We notice that both the classical and the one-instanton
effective action (\ref{seff0}) vanish, so it defines a vacuum. For simplicity we focus on the $U(2)$ case 
and take for the vev
  \be
  v_{Au,Bv}\sim  v_{AB}(\sigma_3)_{u,v}  
  \ee
  The massless degrees of freedom in this branch are described by ${\cal N}=4$ on-shell superfields 
   $\Phi$, $\bar \Lambda$ and ${\cal F}$  proportional to $\sigma_3$.  In the following we restrict ourselves to the effective action for these
   massless degrees of freedom.  Writing $\Phi_{Au,Bv}=\Phi_{AB} (\sigma_3)_{uv} $ and using the algebraic identity 
   \be \label{identity}
       {\rm det}_{[4\times4]} \, 2 \Phi_{AB}  =  (\Phi_{AB} \, \Phi^{AB})^2   \, ,
   \ee
     one finds that the effective action (\ref{seff0}) reduces to a much simpler form
\bea  
S^{k=1}_{\rm eff} =  
c' \, {g^4 \over \pi^6} \,  e^{2\pi \ii \tau} \, \int d^4 a\, d^8 \Theta  \,  {1\over 1-H^{\dot\alpha\dot\beta}\, H_{\dot\alpha\dot\beta}   } 
\label{seff1}
\eea     
with  
\be
H_{\dot\alpha\dot\beta}  = {\mathcal{F}_{\dot\alpha\dot\beta}    +\ft{1}{\sqrt{2}} \bar{\Lambda}_{A\dot\alpha} (\Phi^{-1})^{AB} \bar{\Lambda}_{B\dot\beta}   \over   g\, \Phi^{AB}\Phi_{AB}   }   \, .
\ee

The effective lagrangian can be found by expanding (\ref{seff0}) in $\Theta$ to the order $\Theta^8$ with superfields given by 
 (schematically)  
 \bea
 \Phi &=&  \varphi + \Theta \,  \psi+\Theta^2 F^- \, , \cr
  \bar \Lambda 
&=&   \bar{\psi}+ \Theta  {\partial} \varphi  - 
   \Theta^2   {\partial}  \psi + \Theta^3  {\partial} F^{-}  \, ,\cr
 {\cal F}  &=&  
   F^+   +  \Theta {\partial}  \bar{\psi } + 4 \Theta^2 {\partial}^2 \varphi
+ \Theta^3   {\partial}^2 \psi +
  \Theta^4 {\partial}^2 F^-  \, .
\eea
 Alternatively one can use the twistor variables and write 
\bea
 \Phi_{A B} (x, \Theta)=   e^{\Theta \, \lambda \, \eta } \, \eta_A \,\eta_B \, , 
  \qquad 
  \bar\Lambda_{  \dot{\alpha} A}(x, \Theta) = e^{\Theta \, \lambda \, \eta }  \,\tilde \lambda_{\dot\alpha}   \,\eta_A     \, , \qquad 
 {\cal F}_{\dot\alpha\dot\beta} (x, \Theta)=    e^{\Theta \, \lambda \, \eta }  \,  \tilde \lambda_{\dot\alpha}   \, \tilde \lambda_{\dot\beta} \, .
\eea
 Anti-instantons produce a similar effective action with holomorphic components replaced by anti-holomorphic ones, namely $ F^+ \leftrightarrow F^- $, $ \psi \leftrightarrow \bar\psi $ and $\tau  \leftrightarrow \bar\tau$. With the supersymmetric effective action at hand, one can expand it to the order $\Theta^8$ to obtain one-instanton corrections to the higher derivative terms of interest. In this section, we will focus on the one-instanton effects on $F^n$ and $D^4F^n$, and other higher-dimensional operators can be obtained similarly.  

\subsection{Higher derivative $F^n$ terms} \label{section:Fn}

Here we consider higher derivative terms involving only $F^\pm$ gauge fields, namely $F^n$ terms. The effective action can be read off from (\ref{seff1}) with
 the ${\cal N}=4$ superfields taken to be
 \be
 \bar \Lambda=0 \, ,  \qquad  ~~~~~~~~~ {\cal F}_{\dot\alpha \dot\beta}=F^+_{\dot\alpha \dot\beta}  \, ,  \qquad ~~~~~~  
 \Phi_{AB}=v_{AB}+ \delta \Phi_{AB}   \, ,
 \ee  
 with the fluctuation $\delta \Phi_{AB}$ given by
 \be
 \delta\Phi_{AB}=\ft12\, F^-_{\alpha \beta} \Theta^\alpha_A\, \Theta_B^\beta \, .
 \ee
The effective action reduces to
 \bea
 S_{\rm F^4_-F^{2n}_+}&=& c' \, {g^{4-2n} \over \pi^6} \, e^{2\pi \ii \tau}\, \sum_{n=1}^\infty \, \int  d^4 a\, d^8 \Theta   \,  {  (F^+)^{2n}   \over  (\Phi_{AB} \Phi^{AB} )^{2n}  }  \nn\\
 &=&  
c' \, {g^{4-2n} \over \pi^6} \, e^{2\pi \ii \tau}\, \sum_{n=1}^\infty \, \int  d^4 a\, d^8 \Theta   \,    (F^+)^{2n}    f_8( \delta\Phi,v) \, , \label{sfn2}
 \eea
 where $f_8( \delta\Phi,v)$ comes from expansion of the denominator
 \be
 \label{sfn3}
  f_8( \delta\Phi,v) =\ft{2n(2n+1)}{ v^{4n+4}  }  \left(    \delta\Phi^4-\ft{12 (2n+2)}{v^2} \,  \delta\Phi^2 \, (\delta\Phi \cdot v)^2   +\ft{16 (2n+2)(2n+3)}{ v^8}
 ( \delta\Phi \cdot v)^4\right) \, ,
 \ee
with $v^2=v_{AB} v^{AB}$, $\delta\Phi^2= \delta\Phi_{AB} \delta\Phi^{AB}  $ and  $\delta\Phi \cdot v= \delta\Phi_{AB} v^{AB}$.  
 Integrating over $\Theta$ one finds an infinite series of $(F^-)^4(F^+)^{2n}$ terms with $n\geq 1$. For the special case $n=1$, we find that actually the three terms inside the brackets in (\ref{sfn3}) exactly cancel  against each other upon $\Theta$-integration!  The same conclusions can be reached by considering 
 one anti-instanton corrections to the coupling $(F^-)^{2n} (F^+)^4$. 
 We thus conclude that MHV and $\overline{\rm MHV}$ operators  involving only $F$ fields (and no their derivatives) do not receive one-instanton corrections.  
 This is consistent with the known fact that these operators receive only perturbative corrections at $n$ loops due to the tight constraints from $\mathcal{N}=4$ supersymmetry~\cite{Chen:2015hpa}.  Our results  confirm that this non-renormalization theorem continue to hold at the non-perturbative level! 
 On the other hand for $n\geq 2$ such terms as $(F^+)^{2n} (F^-)^4$ are expected to receive  perturbative corrections already at one loop order and here we show that they also get non-trivial corrections from instantons.

\subsection{$D^4F^n$ terms}

In this section we study one-instanton corrections to MHV higher derivative terms: $D^4F^n$. 
It is easy to check that only the terms built entirely with the superfields ${\cal F}$'s in the effective lagrangian (\ref{seff1}) saturate the MHV bound
  \be
 n_{\bar\lambda} +  2 n_\phi + 3 n_\lambda +  4 n_{F^-} = 8 \, ,
   \ee
   and leads to the coupling $D^4F_-^2F_+^{2p}$. So we set the physical on-shell super fields as
 \be
\bar \Lambda=0 \, , \qquad  ~~~~~~~~~ \Phi_{AB}=v_{AB}   \, ,
\ee
while keeping the superfield ${\cal F}$ general. Plugging this into (\ref{seff1}) one finds the single-trace effective lagrangian   
 \be
 S_{\rm D^4F^{2n}}= c' \, {g^{4-2n} \over \pi^6}  \, e^{2\pi \ii \tau} \, \sum_{n=1}^\infty \, \int  d^4 a\, d^8 \Theta   \,  {  {\cal  F}^{2n}   \over  v^{4n} }  \,,    \label{sfn}
 \ee
 with   $v^2=v_{AB} v^{AB}$ and 
\bea
 {\cal F}_{\dot\alpha \dot\beta}  &=&  e^{\Theta \, \lambda \, \eta }  \,   \tilde \lambda_{\dot\alpha}\tilde \lambda_{\dot\beta}
 =  F^+   +  \Theta {\partial}  \bar{\psi } + 4 \Theta^2 {\partial}^2 \varphi
+ \Theta^3   {\partial}^2 \psi +
  \Theta^4 {\partial}^2 F^-  \, .    \label{ftt}
  \eea
     Performing the integral over $\Theta$  we find the corresponding (super) S-matrix,
     \bea
 S_{\rm D^4F^{2n}} &=& c' \, {g^{4-2n} \over \pi^6} \,  e^{2\pi \ii \tau}\,  \sum_{n=2}^\infty  \delta^{(8)}\left(\sum_{j=1}^{2n} \eta^A_{j}\, \lambda^\alpha_{j} \right) \sum_{\rm perm} \,   {\prod_{i=1}^{n}  \left[ 2i{-}1,2i\right]^2 \,  \over  v^{4n} }  \,,
  \label{sfn}
 \eea
 it is easy to see that the above result has all the correct properties of the (super) S-matrix corresponding to the MHV coupling $(D^2 F^-)^2\,(F^+)^{2n-2}$. Different field components arise from the 
     different powers of $\eta^A_i$  in the expansion of  (\ref{sfn}) or equivalently from the different ways of absorbing the eight $\Theta$ zero modes using the various field components  in the right hand side of  (\ref{ftt}). For example taking from the product only terms containing $\eta^A_{1}$ and $\eta^A_{2}$ 
    one finds 
     \be
 \langle 1 2\rangle^4 \, \sum_{\rm perm}\, \prod_{i=1}^{n}  \left[ 2i{-}1,2i\right]^2: \qquad ~~~~~~ (D^2 F^-)^2\,   (F^+)^{n-2} 
\ee
that corresponds to absorbing the eight fermionic zero modes with  $(D^2 F^-)^2 $ terms.

\section{Perturbation corrections to $D^nF^4$ terms} \label{section:D^nF^4}

In the last section we have shown that operators $D^4 F^2_-F_+^{2n}$ are generally corrected by instantons. In this section we focus on the lowest term $D^4 F^{4}$, and
compute its one and two-loop corrections. The computation here can be easily extended to higher derivative couplings $D^n F^{4}$, and we present the explicit results at the end of this section.

\subsection{Perturbation corrections to $D^4F^4$}

\subsubsection{One-loop contribution to $D^4F^4$} 
 
It is known that four-point amplitudes of $\mathcal{N}=4$ SYM at any loop can be expressed in terms of the tree-level amplitudes multiplied by some loop integrals that form basis \cite{Green:1982sw,Bern:1997nh,Bern:2008pv,Bern:2010tq, Bern:2012uc}. The basis  of one and two-loop integrals for the amplitudes in the Coulomb branch, can be obtained from the massless integrals of $\mathcal{N}=4$ SYM at the origin of moduli space by appropriately replacing original massless propagators with massive W-bosons and unbroken massless gluons~\cite{Alday:2009zm}~\footnote{Recently higher-dimensional operators of six-dimensional supersymmetric Yang-Mills theory in the Coulomb branch were studied in \cite{Chang:2014jta, Lin:2015zea}.}. To be precise, here we are considering the Coulomb branch with the gauge symmetry breaking $U(N+1) \rightarrow U(N) \times U(1)$. At one loop the integral basis comprises scalar box integrals, and the four-point amplitude is given by 
\be 
A^{\rm 1-loop}(1,2,3,4) = {g}^2 N  s_{12}s_{14} A^{\rm tree} (1,2,3,4) \left[ I^{\rm 1-loop}(1,2,3,4) + (2 \leftrightarrow 3 )+ (3 \leftrightarrow 4) \right] \, ,
\ee
where we sum over all independent permutations since we consider the Abelian theory. In the Coulomb branch the internal propagators of the box integrals are massive and the basic integral can be written as
\be 
I^{\rm 1-loop}(1,2,3,4) =-{i  \over (2 \pi)^4}   \int  {d^4 \ell  \over (\ell^2 - m^2) ( (\ell+ k_2)^2 - m^2) ( (\ell+ k_{23})^2 - m^2) ( (\ell- k_1)^2 - m^2) } \, ,
\ee 
with $m={g} \, v$ and $k_{23} = k_2 + k_3$. On the other hand the four-point tree-level amplitude is given by 
\beq
A^{\rm tree} (1,2,3,4) = {g}^2 { \delta^{(8)} (\sum^4_{i=1}\lambda_{i} \eta_{i} ) \over  \langle 12 \rangle \langle 23 \rangle \langle 34 \rangle \langle 41 \rangle  }\, ,
\eeq 
and thus
\be 
s_{12}s_{14} A^{\rm tree} (1,2,3,4) = -{g}^2 \delta^{(8)} (\sum^4_{i=1}\lambda_{i} \eta_{i}) {  [12]^2 \over  \langle 34 \rangle^2  } \, .   \label{atree}
\ee
Not surprisingly we see that the result is precisely the superamplitude generated from the SUSY completion of the coupling 
$F_+^2 F_-^2$, whose S-matrix is simply $[12]^2   \langle 34 \rangle^2$. 
Although not obvious, the result (\ref{atree}) is actually local and permutation invariant because of the four-point restricted kinematics, and the apparent pole at $\langle 34 \rangle = 0$ is an artifact. 

Let us now evaluate the integrals in the large mass expansion. Such integrals, as well as higher-loop ones, for light-by-light scattering have been recently studied in details in~\cite{Caron-Huot:2014lda}. For our purposes of large mass expansion,  the Feynman-parameter representation of the integrals are actually very convenient. 
Using Feynman parametrization, we find that the one-loop integral can be expressed as~\footnote{  Here we use the Feynman  parametrisation of the integrand 
\be 
\prod^n_{i=1} {1 \over A_i} 
= \Gamma(n)  
 { \int d^n \alpha \,  \delta(\sum^n_{i=1} \alpha_i -1 )   \over \left( \sum^n_{i=1} \alpha_i A_i \right)^{n} } \, ,
\ee
and perform the integral with the help of  
\be 
\int  {d^4 \ell  \over (\ell^2 - A)^{n} } =  i \pi^2 { \Gamma(n -2 ) \over \Gamma(n ) \,A^{n-2 } } \, ,
\ee
 }
\be 
I^{\rm 1-loop}(1,2,3,4) = {1\over (4\pi)^2}  \int^{\infty }_0   {  \delta(\sum^4_{i=1} \alpha_i -1 )  \, d^4 \alpha \over (m^2- 
  \alpha_2 \alpha_4 \,s   -\alpha_1 \alpha_3 \,t )^2} \, ,
\ee
Here and in what follows we use the Mandelstan variables $s=s_{12} $,  $ t=s_{23}$ and $u=s_{13}$. 
The integration over $\alpha_i$'s may be further simplified by performing the change of variables, 
\be 
\alpha_1 = \beta_1 \beta_2 \, , \alpha_2 = (1-\beta_1) \beta_3 \, ,
\alpha_3 = \beta_1 (1-\beta_2) \, , \alpha_4 = (1-\beta_1) (1-\beta_3) \, ,
\ee
with a Jacobian factor $||\partial\alpha/\partial\beta|| = \beta_1 (1 - \beta_1)$. The new variables automatically solve the constraint $\sum^4_{i=1} \alpha_i =1$, too, and lead to the integral
\be 
I^{\rm 1-loop}(1,2,3,4) =  {1\over (4\pi)^2}  \int^{1 }_0   { d^3 \beta \,  \beta_1 (1 - \beta_1)\over (m^2- 
  (1-\beta_1)^2 \beta_3  (1-\beta_3) \,s  - \beta^2_1 \beta_2  (1-\beta_2) \, t )^2} \, .
\ee
With the expression in terms of Feynman parameters at hand, one can simply expand the integrand in the large mass limit, which then leads to simple integrations over polynomials. In this way one can obtain $D^n F_-^2 F_+^2$ for very high $n$ quite easily. To leading order, namely for $n=0$, it gives 
\be 
I^{\rm 1-loop}_0(1,2,3,4) =  {1\over (4\pi)^2}  \int^{1 }_0 
{d^3 \beta \, \beta_1 (1 - \beta_1) \over  m^4 }  =  {1\over 6 (4\pi)^2 \,m^4}     \, .
\ee 
Thus summing over the three permutations, we find the term $F_-^2 F_+^2$ generated at one loop with  coefficient 
\be
F_-^2 F_+^2 :  \qquad ~~~~~~~~~~~~~~ { {g}^4 N \over 2 (4 \pi)^2 \, m^4 } \, .   \label{f4}
\ee
At the first sub-leading order, the large mass expansion leads to a result proportional to $s+t+u$ corresponding to the operator $D^2 F_-^2 F_+^2$, which vanishes with on-shell condition, namely $s+t+u=0$. At the next order one finds
\beq
I^{\rm 1-loop}_4(1,2,3,4) = 
 { 2 s^2 + s\,  t + 2 t^2  \over 840 (4 \pi)^2 m^8 } \, .
\eeq
Summing over permutations leads to $ {s^2+t^2+u^2 \over  240 (4 \pi)^2  m^8 } $. Thus the one-loop contribution to the 
coefficient of the MHV coupling $D^4F_-^2 F_+^2$ is given as
\beq
D^4F_-^2 F_+^2  :  \qquad ~~~~~~~~~~~~~~ {{g}^4 N \over  240 (4 \pi)^2  m^8 }  \, .\label{oneloopd4f4}
\eeq

\subsubsection{Two-loop contribution to $D^4F^4$} 

Let us now consider the same amplitude at two loops 
\beq \label{eq:twoloop}
A^{\rm 2-loop}_4 = \left( {g}^2 N \right)^2 (1+ {1 \over N}) s_{12}s_{14}A^{\rm tree}_4 \left[ s_{12} (I^{\rm 2-loop}(1,2,3,4)+I^{\rm 2-loop}(3,4,2,1)) +{\rm cyclic(2,3,4)} 
\right] \, ,
\eeq
where the factor $(1+ {1 \over N})$ is due to the fact that the internal massless propagator of the scalar integral $I^{\rm 2-loop}(1,2,3,4)$ shown below can be photon or gluon. The scalar integral $I^{\rm 2-loop}(1,2,3,4)$  
\beq \nonumber
 \includegraphics[scale=0.4]{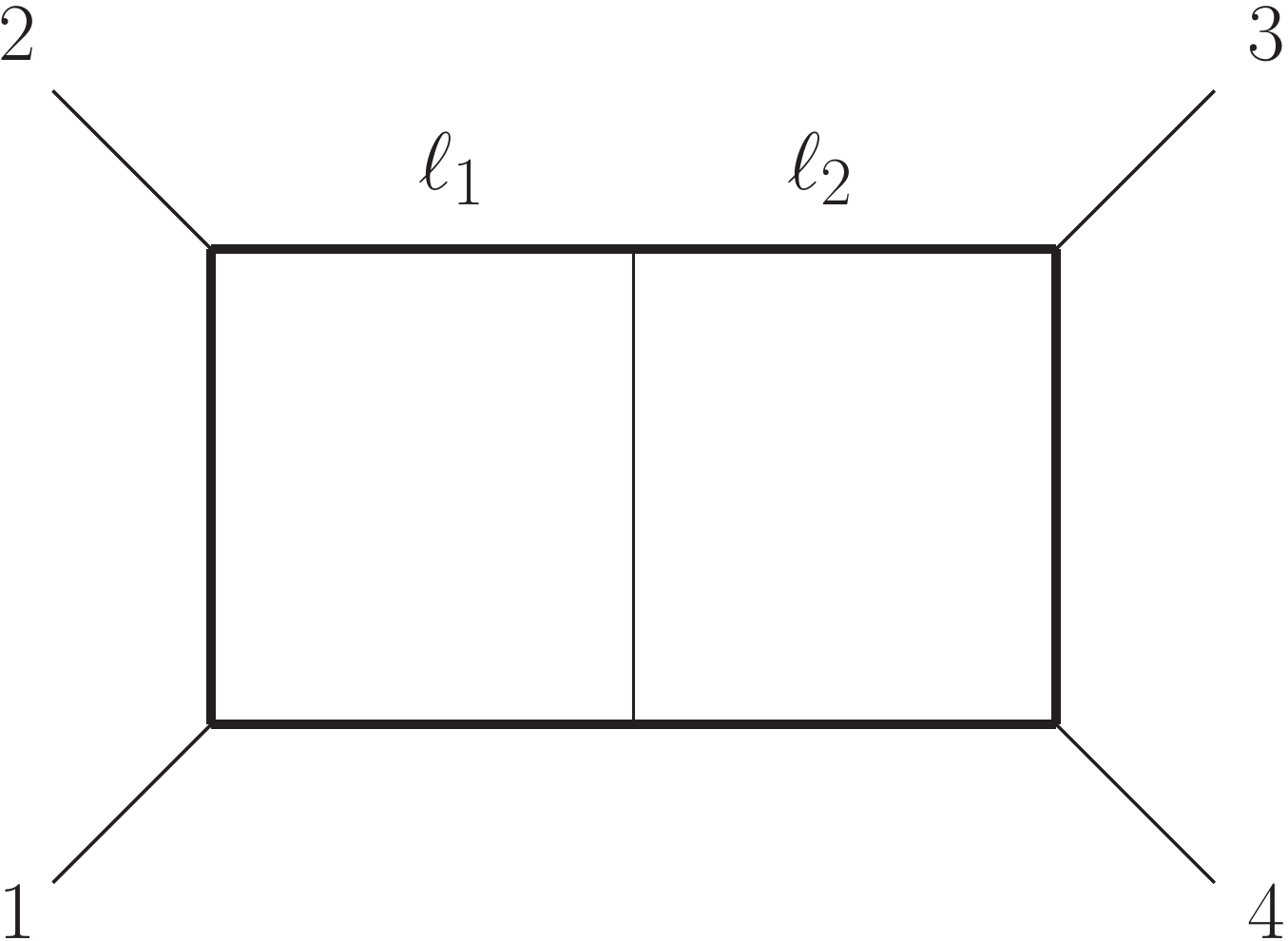}
\eeq
is defined by 
\beqa \label{eq:twoloop2}
&& I^{\rm 2-loop}(1,2,3,4)
= - \int {d^4 \ell_1 d^4 \ell_2 \over (2 \pi)^8} {1 \over ( \ell_1^2 -m^2 )[(\ell_1 + k_2)^2 -m^2 ][(\ell_1 + k_{23})^2 - m^2]} \cr
&&~~~~~~~~~~~~~~\times 
{1 \over ( \ell_2^2 -m^2 )[(\ell_2 + k_3)^2 -m^2][(\ell_2 + k_{34})^2 - m^2]( \ell_1 + \ell_2 )^2 } \, .
\eeqa
Here we have omitted the following non-planar contributions, 
\beq \nonumber
 \includegraphics[scale=0.4]{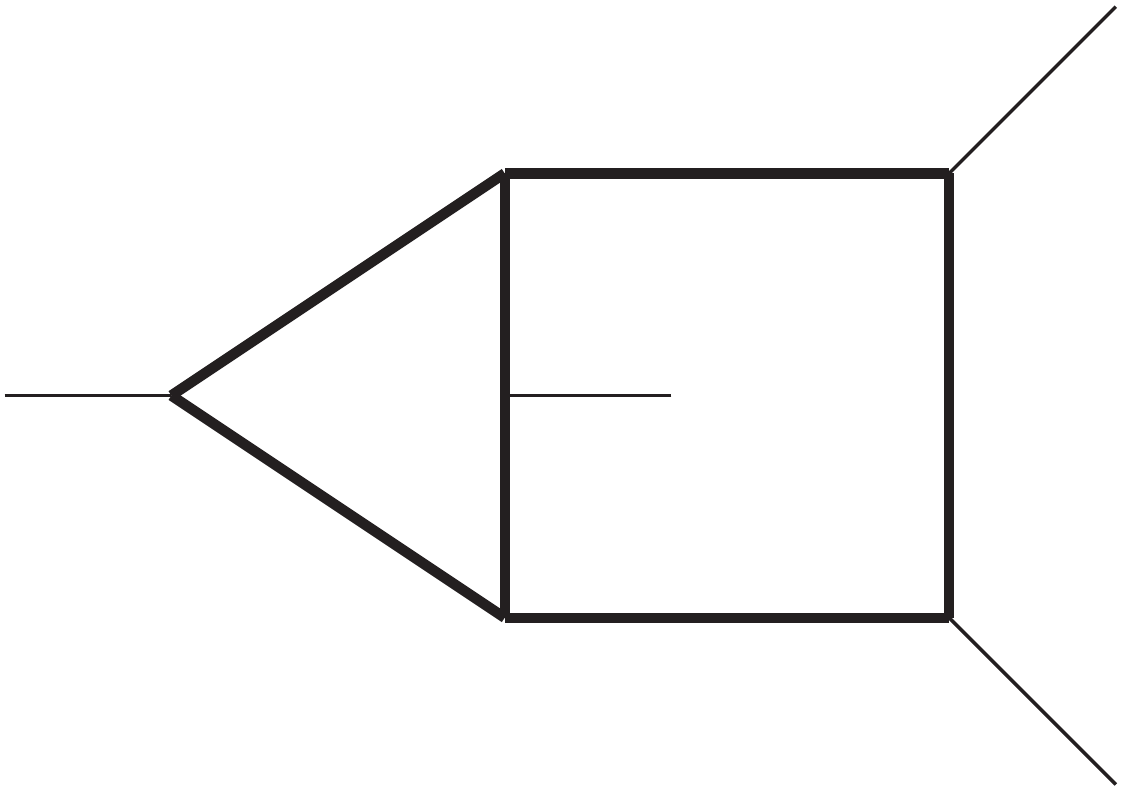}
\eeq
that is because taking massless (mutually Abelian) external lines  forces all the internal propagators to be massive (charged) W-bosons, denoted in the above diagram by thick lines. However there is no three massive W-boson vertex. 

From (\ref{eq:twoloop}) and (\ref{eq:twoloop2}), it is easy to see that the leading order in the large-mass expansion corresponds to the coupling $D^2F_-^2F_+^2$, which again vanishes on-shell. This implies that the $F^2_-F^2_+$ coupling does not receive two-loop corrections, in agreement with the non-renormalisation theorem~\cite{Dine:1997nq}, which states that the coefficient of $F^2_-F^2_+$ is in fact one-loop exact. 
The first non-trivial correction appears at order $D^4F_-^2F_+^2$.

To evaluate the integral $I^{\rm 2-loop}(1,2,3,4)$, we again express the integral in terms of Feynman parameters  so that it reads
\beq
I^{\rm 2-loop}(1,2,3,4) =  \Gamma(3) \int^{\infty }_{0}  d^7 \alpha \, \delta(\sum^7_{i=1} \alpha_i -1 ) \, { \mathcal{U}  \over \mathcal{V}^3 } \, ,
\eeq
with $ \mathcal{U}$ and $ \mathcal{V}$ given by
\beqa
 \mathcal{U} &=& ( \sum^4_{i=1} \alpha_i )( \sum^7_{i=4} \alpha_i )- \alpha^2_4 \, , \\
 \mathcal{V} &=& \mathcal{U} ( \sum_{i \neq 4} \alpha_i )  m^2  - \alpha_1 \alpha_4 \alpha_7 s_{41} - \left[ \alpha_2 \alpha_3 (\sum^7_{i=4} \alpha_i  ) + \alpha_5 \alpha_6 (\sum^4_{i=1} \alpha_i  ) 
 + \alpha_4 ( \alpha_2 \alpha_6 + \alpha_3 \alpha_5 ) \right] s_{12} \,  .\nonumber 
\eeqa
The integration over $\alpha_i$'s can be further simplified using the change of variables~\cite{Smirnov:1999gc}  
\bea \label{changevariable}
\alpha_1 &=&  \beta_4 (1 - \beta_5) \, \beta_3 \,\quad   \alpha_2 = (1 - \beta_4 )(1 - \beta_5 ) \,   \quad   \alpha_3 =  \beta_4 (1 - \beta_5) \,  (1-\beta_3) \,  \nn\\
\alpha_5 &=& \beta_2\beta_5 (1-\beta_1) \,\quad  \alpha_6 = \beta_5 (1 - \beta_2 ) \, 
\quad   ~~~~~~~~\alpha_7 = \beta_2\beta_5 \beta_1 
\eea
with a Jacobian factor of $\beta_2 \beta_4 \beta^2_5(1-\beta_5)^2$.  Furthermore, one can rely on the so-called Cheng-Wu theorem~\cite{Cheng-Wu} which allows one 
to replace the delta function by  $\delta( \sum^7_{i =1 } \alpha_i -1 )$ with the sum now running 
over an arbitrary subset ${\cal M}$ of  $\{1, 2, \ldots, 7\}$ without changing the result. After the replacement the integration regions of the variables in the subset ${\cal M}$
 can be reduced to $[0, 1]$ due to the delta-function constraint. The most convenient choice in our case is ${\cal M} = \{1, 2, 3,5,6, 7\}$ and the Feynman parametrisation of the integral can be
 written as
 \beq
I^{\rm 2-loop}(1,2,3,4) =  \Gamma(3)   \int^{\infty }_{0} d \alpha_4 \int^1_0 d^5 \beta \, \beta_2 \beta_4 \beta^2_5(1-\beta_5)^2 \,    { \mathcal{U}  \over \mathcal{V}^3 } \, ,
\eeq
After expanding in the large mass limit, the best way of performing the integral is to carry out the $\alpha_4$ integration first. One then is left with a polynomial in $\beta_i$'s, whose integral can be evaluated efficiently.

As observed earlier, the first non-trivial case is the coupling $D^4F^2_-F^2_+$. Performing the integrations over Feynman parameters for $I^{\rm 2-loop}(1,2,3,4)$ at this order leads to the result 
\beq
I^{\rm 2-loop}_4(1,2,3,4) = { 8 s + t \over 360 (4 \pi)^4  m^8 } \, ,
\eeq
Plugging this into (\ref{eq:twoloop}), one finds
\be
 s_{12} (I^{\rm 2-loop}_4(1,2,3,4)+I^{\rm 2-loop}_4(3,4,2,1)) +{\rm cyclic(2,3,4)} =
   {1 \over 24 (4 \pi)^4 m^8} (s^2 +t^2 + u^2) \,.  
 \ee
Namely, the two-loop contribution to the coefficient of  the operator $D^4F^2_-F^2_+$ is 
\beq
D^4F^2_-F^2_+ : \quad ~~~~~~~~~~~~~~~~~~ { {g}^6 N^2 \over 24 (4 \pi)^4 m^8} (1 + {1 \over N}) \,.  \label{twoloopd4f4}
\eeq

\subsubsection{One- and two-loop contributions to $D^nF^4$}

The formulae in the last section can be easily expanded to higher orders in the limit of large mass to extract higher derivative terms $D^{n}F_-^2 F_+^2$. 
 It is known that from mass dimension $0$ to $10$ there is only one independent kinematic structure for each order, namely $s^n+t^n+u^n$, while starting from dimension $12$ there will be more than one independent kinematic structures (except at dimension $14$ where there is again only one structure, $s^7+t^7+u^7$)~\cite{Elvang:2010jv, Boels:2013jua}.
 In table \ref{tab:onetwoloop} we display the results we find for $n \leq 10$ by explicit evaluation of the integrals at one and two loops.
 \begin{table}[!h]
\centering 
\begin{tabular}{c|ccccccccccccccc}
n &  0 & 2 & 4 & 6 & 8 & 10 
 \\ \hline
{one-loop} &  ${1 \over 2}$ & $0$ & ${1 \over 240}$ & ${1 \over 1512}$ & ${1 \over 7560}$ & ${1 \over 39600}$ 
  \\
{two-loop} &  $0$ & $0$ & ${1 \over 24}$ & ${7 \over 720 }$ & ${47 \over 20160}$ & ${7319 \over 13608000 }$
\end{tabular}
\caption{A list of results for $D^nF_-^2 F_+^2$ at one loop. Here the first row refers to the mass dimension, and we have omitted an overall factor of ${{g}^4 N \over (4 \pi)^2 m^{4+n}}$ and ${ {g}^6 N^2 \over (4 \pi)^4 m^{4+n}} (1 + {1 \over N})$ for one- and two-loop results respectively.
\label{tab:onetwoloop}}
\end{table}
On the other hand at $n=12$ we find for the two different structures are given by 
\bea 
{\rm one-loop} :&&  {{g}^4 N \over (4 \pi)^2 m^{16}} {1 \over 192192}  \left [ (s^6+t^6+u^6)-{ 7 \over 75} (s^2 \, t^2\,u^2)  \right] \\
{\rm two-loops} :&&  {{g}^6 N^2 \over (4 \pi)^4 m^{16}} (1 + {1 \over N}) {7541 \over 59875200 } \left [ (s^6 + t^6 +u^6)-{ 249 \over 7541} (s^2 \, t^2\,u^2)  \right] \, . \nn
\eea
We have further checked that upto $n=20$ that very non-trivial kinematics structures appear at both one and two loops, and as the case of $n=12$ discussed above they are all different at different loop orders.


\subsection{ $F^n$ terms }

 \subsubsection{The one-loop corrections}

The one-loop contribution to $F^n$ terms has been computed in~\cite{Buchbinder:2001ui}.
The result arises from the one-loop determinant of an open string
in the background of a constant field $F$. The result is expressed in terms of skew-diagonalized $F_{mn}$ with the choice, 
\be
F_{12}=f_1  \, ,   \qquad ~~~~~~~~    F_{34}=f_2 \, .
\ee  
The one-loop determinant then reads~\cite{Buchbinder:2001ui}
\bea
S_{\rm one-loop} &=& { 8 N \over (4 \pi)^2}   \, \int_0^\infty {d{\sigma} e^{-{\sigma} v^2 {g}^2 } \over  {\sigma}^3} {g f_1 {\sigma}\over {\rm sinh} (g f_1 {\sigma}) }  \, {g f_2 {\sigma}\over  {\rm sinh} (g f_2 {\sigma}) } (   {\rm cosh} (g f_1 {\sigma}) - {\rm cosh} (g f_2 {\sigma})   )^2\nn\\
 &=& {N \over (4 \pi)^2} \left(  { F_+^2 \, F_-^2  \over 2 v^4 } +
 { F_+^4 \, F_-^4   \over 16 v^{12} {g}^4 }    - { 5(F_+^6 \, F_-^4+ F_+^4 \, F_-^6)   \over 96 \, v^{16} {g}^{6} }  +\ldots \right) \, , \label{tseytlin}
\eea
where on the second line we have written the final result in terms of Lorentz invariant quantities by using 
\bea
f_1^2 + f_2^2 = {1 \over 4} (F_+^2 + F_-^2) \, , \quad 
f_1  f_2 = {1 \over 8} (F_+^2 - F_-^2) \, .   \label{fFF}
\eea 
We note for the MHV terms $F_-^2 F_+^{2n}$ only the lowest operator $F_+^2 F_-^2$ appears, while all the higher-dimensional operators are absent as it should be since $F_-^2 F_+^{2n}$ can only generate perturbatively at $n$ loops as we remarked earlier. Finally all the non-MHV operators such as $F_-^4 F_+^{2n}$ already appear at one loop, and expected to receive higher-loop as well as instanton corrections, which is indeed the case from our one-instanton results, in particular the results in section \ref{section:Fn}.

\subsubsection{The HEA description}

As already noticed and commented in~\cite{Schwarz:2013wra}, only MHV $F^n$ terms  are correctly reproduced by the DBI action
describing the  gauge theory on a single D3-brane in the AdS$_5 \times S^5$ background ( the HEA in  \cite{Schwarz:2013wra, Schwarz:2015lla} ).
Indeed, expanding this action for gluon fluctuations one finds 
\bea \label{eq:DBI}
{\cal L}_{\rm DBI} &=& {1\over   \kappa^2}  \left(  1 - \sqrt{{\rm det}(1+{\kappa} F)} \right) =
{1\over   \kappa^2}  \left(  1 - \sqrt{ (1+{\kappa}^2 f_1^2) (1+{\kappa}^2 f_2^2)  } \right) 
\cr
&=& - {1 \over 8} (F^2_+  + F_-^2)+ { \kappa^2 \over 2 } (F^2_- F^2_+)
-    \kappa^4  ( F^2_- F^4_+ + F^2_+ F^4_- ) \cr
&{}&
+2 \,    \kappa^6 \,  ( F^2_- F^6_+ + F^2_+ F^6_- + 3 F^4_- F^4_+ ) + \ldots \, ,
\eea
 where (\ref{fFF}) have been used to write the determinant in terms of Lorentz invariants, 
 \be
 {\kappa}^2=  {   N \over   4 \pi^2 v^4 } \, ,
 \ee
 and $v$ is the scalar vev. 
 We note that the $F_+^2 F_-^2$ term matches the perturbative result in (\ref{tseytlin}). Indeed the HEA gives correct coefficients for all MHV terms $F_-^2 F_+^{2p}$ (as well as their conjugates) consistently with the non-renormalization theorem of these terms. 
  Non-MHV terms such $F_+^m F_-^n$ with both $m,n$ larger than 4 receives both perturbative and instanton corrections. This is consistent
  with S-duality, since the equations of motion following from (\ref{eq:DBI}) have been shown to be  
  $SL(2,\Z)$-invariant, so much so that any perturbative correction
  to this action should be accompanied by instantons in an $SL(2,\Z)$-invariant manner. More precisely, taking for simplicity $\tau_1=0$ and choosing skew-diagonalized $F_{mn}$ with $F_{12} =f_1$ and $F_{34} = f_2$, 
   S-duality sends  $S:~ \tau\to -{1\over \tau}$ and exchanges $f_i$ and its dual  $h_i=- \ii  \partial  {\cal L}_{\rm DBI} /\partial f_i$. 
   For the DBI Lagrangian (\ref{eq:DBI}) one finds \cite{Schwarz:2013wra, Schwarz:2015lla}\footnote{We work on the Euclidean signature and in the field theory
   basis (see above). } 
  \be
h_1 =-\ii f_1 \sqrt{ 1+\kappa^2 \, f_1^2 \over  1+\kappa^2 \, f_2^2     } \, , \quad \quad ~~~~~  
h_2 =-\ii f_2 \sqrt{ 1+\kappa^2 \, f_2^2 \over  1+\kappa^2 \, f_1^2     }  \, ,     \label{hfdual}
\ee 
 that is solved by identical formulae for $f_i$ in terms of $h_i$ with $f_i \leftrightarrow h_i$.   This shows the  symmetry between the $f_i$ and $h_i$ S-dual descriptions. The above duality transformation would be changed if the action 
 were modified by adding corrections we discussed in the paper, however, by considering small $f_i$ expansion, $f_i \to h_i= -\ii f_i+\ldots ,$ reduces to the linear action (\ref{fdual}).  In the next section
 we will limit ourselves to the linear definition of S-duality, from which one can already ask whether $SL(2,\Z)$ completed results are consistent with the one-instanton effective action we obtained. 
 
 \section{$SL(2,\Z)$ completion} \label{section:SL(2,Z)}
 
  The effective actions of ${\cal N}=4$ SYM and of the Type IIB D3-brane have been conjectured to enjoy an exact strong-weak coupling $SL(2,\Z)$ symmetry. 
  The requirement of $SL(2,\Z)$ symmetry provides very tight constraints on the possible form of higher dimensional terms in the effective action and
  on its dependence on the gauge coupling constant. This requirement allows often to determine the exact form of the 
  coupling \cite{Green:1997di,Green:2000ke,Green:2005ba} starting from a few perturbative data. 
  In \cite{Green:2000ke},  string corrections to the  $D^4 F^4$ couplings in the world volume of a single D3-brane were considered. This coupling 
  is expected to receive perturbative corrections only at tree level and one loop. An $SL(2,\Z)$-invariant formula was derived by starting from the tree-level and one-loop
  contributions (proportional to $\tau_2$ and $\ln(\tau_2)$, respectively) and summing up over all its $SL(2,\Z)$ images. This results into the modular invariant non-holomorphic
  function \cite{Green:2000ke}
   \be \label{eq:Z1}
Z_1 = \sum_{m,n\neq 0} {}^{'}{\tau_2 \over |m-n \tau|^{2}    }= \ln |\tau_2 \, \eta(\tau)^4| \,,
\ee
where $\eta(\tau)$ is the Dedekind function, and the ``primed sum" indicates that the summation is regularised by subtracting a logarithmic divergent constant, which has no $\tau$-dependence as can be checked by taking a derivative w.r.t. $\tau$ or $\bar \tau$ and getting a convergent result. This modular function admits the weak coupling expansion in the large $\tau_2$ limit \cite{Green:2000ke}
 \be
     Z_1 = 2\zeta_2 \, \tau_2 -{\pi\over 2} \ln \tau_2+\pi 
     \sum_{k=1}^\infty \sum_{d|k}    \, {1 \over d} \,  \left( e^{2\pi \ii k\,\tau}+  e^{-2\pi \ii k\, \bar\tau} \right) \label{esmod}
    \ee 
 showing an infinite tower of instanton corrections besides tree-level and one-loop contributions. Here $d|k$ indicates that we sum over $d$'s which are the divisors of $k$. 

  Here we consider the same $D^4 F^4$ coupling but in the non-abelian gauge theory (yet in the Coulomb branch).  The role of $\alpha'$ is now played by the 
  vev $v$ of the scalar field. There is however an important difference with respect to the $U(1)$ open string theory in \cite{Green:2000ke}.
   In the case of the non-abelian theory,  there is no obvious reason while the coupling $D^4 F^4$ cannot get higher loop corrections. Indeed, the explicit
   computations in the last section shows that the higher derivative term is certainly corrected at one and two loops.  Still one can proceed in the same way as before and 
   look for the function in front of the higher derivative term as a sum of modular invariant functions accounting for the perturbative 
   contributions  (proportional to $\tau_2^s$ for some $s$) and their $SL(2,\Z)$ images \cite{Green:2005ba}
    \be
   Z_s= \sum_{m,n\neq 0} {\tau_2^s \over |m-n \tau|^{2s}    } \, ,
    \ee
with $s>1$, for the case $s=1$, the above summation is divergent and $Z_1$ is defined previously in (\ref{eq:Z1}). 
This function is modular invariant by construction and  it is easy to see that is an eigenvector of the Laplace operator 
    \be
    \Delta Z_s= 4 \tau_2^2 {\partial^2 \over \partial \tau\partial \bar\tau} Z_s=s(s-1) Z_s \, .
    \ee
 The  weak coupling expansion of $Z_s$ is given by
    \bea
      Z_s &=& 2\zeta_{2s} \, \tau_2^s+2 \sqrt{\pi}\, \zeta_{2s-1} \, {\Gamma(s-\ft12)\over \Gamma(s)}\, \tau_2^{1-s}\nn\\
&&   +{2\pi^s\over \Gamma(s) }  \sum_{k=1}^\infty \sum_{d|k}  {k^{s-1}\over d^{2s-1}}  \,  \left( e^{2\pi \ii k\,\tau}+  e^{-2\pi \ii k\, \bar\tau} \right)  \left(1+ {s(s-1)\over 4\pi  k  \tau_2} +\ldots \right) \label{esmod2}
    \eea
   for $s>1$. 
   For instance from (\ref{oneloopd4f4}) and  (\ref{twoloopd4f4}) , the one and two loops corrections to the 
   $D^4 F^4$ term read
\beq
 f^{D^4 F^4}_{\rm pert} (\tau,\bar \tau) = {   N \over  240 (4 \pi)^4   v^8 }   \left( \tau_2^2 +  { 5 (N+1) \over 2\pi } \,\tau_2         +\ldots \right)  \,.
\eeq
with $\tau_2={4\pi\over  {g}^2 }$ and $m={g}\, v$. Summing over their $SL(2,\Z)$ images one finds
\beq
 f^{D^4 F^4}(\tau,\bar\tau)= {   N \over  240 (4 \pi)^4   v^8 }   \left( {  Z_2 \over 2 \zeta_4 }   +  { 5 (N+1) \over 2\pi }       { Z_1 \over 2 \zeta_2 }     +\ldots \right) \,,   \label{fresults}
\eeq
for the $U(2)$ gauge group, one should take $N=1$ here.  We remark that the sums over $m,n$ in $Z_2$ and $Z_1$ account for the contribution of the infinite tower 
  of dyons (the $SL(2,\Z)$ images of the massive gluon) with mass $M=|m - n \tau| v$ running in the loops. However, it is quite remarkable that the instanton contributions coming from the two series
 are both proportional to $g^0 \pi^{-6}$ and only differ by a numerical (rational) coefficient.  
 This $g^0 \pi^{-6}$ factor precisely matches our one-instanton result in (\ref{sfn}) with $n=2$! This agreement with right powers both of the coupling constant $g$ and $\pi$ is rather non-trivial since from (\ref{seff0}) we see that the $g$-dependence can be very different for different terms in the expansion. Thus this agreement may be taken as a hint to the validity of the $SL(2,\Z)$ completion. On the hand, besides one- and two-loop perturbative as well as instanton non-perturbative results, $Z_1$ and $Z_2$ also contains two peculiar contributions: $\ln(\tau_2)$ and $1/\tau_2$. Like the more familiar instanton contributions they arise as the result of the sum over the infinite tower of dyonic $SL(2,\Z)$ images of the gluon.
 The field theory interpretation of these two terms is far from obvious. Since $1/\tau_2 \sim g^2 \sim g^{10}/m^8$, by counting the power on the coupling constant $g$, one would expect  
 this contribution appear also at four loops. Still the transcendental $\zeta_3$ coefficient weighting this term induces us to be cautious. The interpretation of the  ${g^8\over m^8} \ln(\tau_2)$ 
term is even more challenging. Progress in this direction requires an explicit evaluation of 3- and 4-loop corrections to the $D^4F^4$ term.  
  
  Similar $SL(2,\Z)$ completions can be written down for the $F^n$ terms. For example for the $F_+^4 F_-^4$ one can find the one loop results in (\ref{tseytlin})
  \beq
   f^{F_+^4 F_-^4}_{\rm pert} (\tau,\bar \tau) = {   N \over   (4 \pi)^4   v^{12} }   \tau_2^2      +\ldots 
  \eeq
 leading to 
 \beq
 f^{F_+^4 F_-^4}(\tau,\bar\tau)= {   N \over   (4 \pi)^4   v^{12} }   {  Z_2 \over 2 \zeta_4 }      +\ldots   \,.   \label{fresults2}
\eeq
 so the instanton part is proportional to $g^0\pi^{-6}$ in agreement again with the one-instanton result. Similarly the one-loop result of $F_+^6 F_-^4$ is given as
 \beq
   f^{F_+^6 F_-^4}_{\rm pert} (\tau,\bar \tau) = -{   5N \over  96 (4 \pi)^5   v^{16} }   \tau_2^3      +\ldots 
  \eeq
 This term itself transforms non-trivially under $SL(2,\Z)$-duality:
 \beq
   F_+^6 F_-^4   \to   {  m-n \tau \over  m-n \bar \tau    } \,  F_+^6 F_-^4
 \eeq
   The  summation over the $SL(2,\Z)$ images leads then to
 \beq
    \sum_{m,n\neq 0} {\tau_2^3 \over |m-n \tau|^{6}    }  {  m-n \tau \over  m-n \bar \tau    }={2\ii \tau_2\over 3} \partial_\tau Z_3
 \eeq
 so for the function of the coupling one finds 
 \beq
 f^{F_+^6 F_-^4}(\tau,\bar\tau)= {   5 N \over  96 (4 \pi)^5   v^{16} }   {  \ii \tau_2\, \partial_\tau Z_3 \over  3 \zeta_6 }      +\ldots   \,.   \label{fresults3}
\eeq
  We notice that now the leading instanton  part in (\ref{fresults3}) is proportional to $g^{-2} \pi^{-6}$ in agreement with the one-instanton result.   Higher $F^n$ terms can be analysed in a similar way. We find that the $SL(2,\Z)$ completion of the one-loop results are always consistent with the one-instanton effective action. 
 
Summarizing we find that $SL(2,\Z)$-duality can be used to complete results from perturbation theory and predicts a very precise form for multi-instanton contributions. Waiting for a better knowledge  of the multi-loop perturbative results of higher-dimensional terms, and a systematic study of higher derivative corrections
 to the duality map we check that one-instanton results are compatible
 with the predictions of  $SL(2,\Z)$ invariance.  It is challenging, and would be of great interest  to test these predictions  at a multi-instanton level against microscopic instanton computations based on localisation techniques that have proven to be extremely powerful in ${\cal N}=2$ multi-instanton calculus \cite{Nekrasov:2002qd, Bruzzo:2002xf, Nekrasov:2003rj, Fucito:2009rs, Billo':2010bd}.

\section{Conclusions and outlook} \label{section:conclusion}

In this paper, we initiated a systematic study on the instanton corrections to the effective action of $\mathcal{N}=4$ SYM. We derived the one-instanton effective action with manifest $\mathcal{N}=4$ on-shell supersymmetry. The action is obtained by studying the system of D(-1)-D3 branes, where D(-1)-branes play the role of instantons. Expanding the supersymmetric effective action in components, we obtained explicit one-instanton corrections to higher derivative terms of interest. In particular, we studied in details higher-dimensional terms such as $F^n$ and $D^4F^n$. We confirmed that non-trivially, {\it i.~e.} thanks to remarkable cancellations, the known non-renormalization theorems for the so-called MHV operators $F_-^2 F_+^{2p}$ do hold at the non-perturbative level, while non-MHV terms such as $F_-^2 F_+^{2p}$ with $p>2$ do generally receive instanton corrections. We also computed the one and two-loop perturbation contributions to $D^4F^4$ (more generally $D^mF^4$), and the results are eventually promoted into modular forms by summing over their $SL(2, \Z)$ images. The proposed $SL(2, \Z)$ invariant modular forms are consistent with the perturbation results as well as the non-perturbative one-instanton effective action at both ends of the large $\tau_2$ expansion. The $SL(2, \Z)$ completed result also leads to some non-trivial predictions for the coupling $D^4F^4$. In particular it determines a precise form of multi-instanton contributions. It would be of great interest to verify (or disprove) the $SL(2, \Z)$ prediction by explicit multi-instanton computations, which we will leave as a future research direction. Indeed the localisation techniques in supersymmetric gauge theories have been extremely powerful and fruitful, and may be utilized for the particular problems of our interest.

Higher-loop perturbation results on $D^4F^4$ (as well as other operators) would be another very important data for fully determining its coefficient. Recently there has been tremendous progress in computing and understanding scattering amplitudes in $\mathcal{N}=4$ SYM. However, most of the advance and interest has been focusing on amplitudes at the origin of the moduli space. Here we are interested in the amplitudes on the Coulomb branch, and more precisely their large mass expansion. Although there has been some interesting investigation on the scattering amplitudes on the Coulomb branch of $\mathcal{N}=4$ SYM, for instance~\cite{Alday:2009zm, Craig:2011ws}, most of the focus has been on the small mass limit, which is opposite to the limit we considered here. Exploring this relatively new zone of $\mathcal{N}=4$ SYM may eventually uncover other surprising and beautiful features of the theory. A nice example of such possibility is the previously mentioned non-renormalization theorem on the MHV terms $F_-^2 F_+^{2p}$, which allows one to completely determine the coefficient of $F_-^2 F_+^{2p}$ in terms of that of the lowest operator $F_-^2 F_+^{2}$. It would be very interesting whether such non-renormalization theorems can be further extended to other higher derivative operators, which may require to consider other symmetry principles beyond $\mathcal{N}=4$ supersymmetry.

\section*{Acknowledgement}
We would like to thank Andreas Brandhuber and Gabriele Travaglini for collaboration at an early stage of this project and for helpful correspondence. We would also like to thank Lance Dixon,  Burkhard Eden, Francesco Fucito, Paul Heslop, Yu-tin Huang, Valya Khoze, Stefano Kovacs,  Sergei Kuzenko, John Schwarz, Yassen Stanev, Jan Troost and Arkady Tseytlin for helpful discussions. 


\providecommand{\href}[2]{#2}\begingroup\raggedright
 \bibliography{referencesEA}
 \bibliographystyle{abe}

\end{document}